\documentclass[11pt,twoside]{article}

\usepackage[utf8]{inputenc}
\usepackage{epsfig,amsfonts,color}
\usepackage{amsmath, bbm, mathabx,mathrsfs}
\usepackage[normalem]{ulem}
\usepackage{multirow}
\usepackage{amssymb, palatino, geometry,url}
\usepackage[colorlinks=true,linkcolor=blue,citecolor=blue,urlcolor=blue]{hyperref}
\usepackage{subcaption}
\usepackage{multirow}
\usepackage{booktabs}
\usepackage{adjustbox}
\usepackage[title]{appendix}
\usepackage{tabularx}
\usepackage{optidef}
\usepackage{graphicx}
\usepackage{forest}
\usepackage{tikz-qtree}
\usepackage{algorithm}
\usepackage{algpseudocode}
\usepackage{makecell}
\usepackage{nicematrix}

\usepackage[dvipsnames]{xcolor}
\usepackage{accents}
\usepackage{listings}
\DeclareCaptionFont{white}{\color{white}}
\DeclareCaptionFormat{listing}{{\parbox{\dimexpr\textwidth-2\fboxsep\relax}{#1#2#3}}}
\lstdefinelanguage{julia}
{
keywordsprefix=\@,
morekeywords={
exit,whos,edit,load,is,isa,isequal,typeof,tuple,ntuple,uid,hash,finalizer,convert,promote,
subtype,typemin,typemax,realmin,realmax,sizeof,eps,promote_type,method_exists,applicable,
invoke,dlopen,dlsym,system,error,throw,assert,new,Inf,Nan,pi,im,begin,while,for,in,return,
break,continue,macro,quote,let,if,elseif,else,try,catch,end,bitstype,ccall,do,using,module,
import,export,importall,baremodule,immutable,local,global,const,Bool,Int,Int8,Int16,Int32,
Int64,Uint,Uint8,Uint16,Uint32,Uint64,Float32,Float64,Complex64,Complex128,Any,Nothing,None,
function,type,typealias,abstract,get_node,create_estimation_model,set_solution, Matrix, Dict,
solve, get_solution, solve_ss_problem, create_estimation_problem, addnode, nothing, 
set_optimizer_attribute, optimize!, println, value, termination_status, objective_value, Model,
optimizer_with_attributes, zeros, length, getJuMPmodel, savefig, plot, plot!, scatter!, grid, 
xlabel!, ylabel!, xlabel, ylabel, zip, sum, InfiniteModel, seed!, Normal, mean, var, Infinite,
OrthogonalCollocation, collect, OptiGraph, set_optimizer, eachindex, set_parameter_value, 
Parameter, dual, set_silent, InfiniteLogical, Disjunct, Hull, InfiniteGDPModel, ExaTranscriptionBackend, CUDABackend, simulate, allowscalar, warmstart_backend_start_values, sin
},
morekeywords = [2]{},
alsoletter=!,
sensitive=true,
morecomment=[l]{\#},
morestring=[b]"
}

\newcommand*{\topbordermatrix}[2]{%
    {\settowidth{\dimen0}{$\scriptstyle\text{#1}$}\kern\dimen0}%
    {\settowidth{\dimen0}{$A$}\kern-\dimen0}%
    \kern-\tabcolsep
    \bordermatrix{\hfill\llap{$\scriptstyle\text{#1}$}\kern-\tabcolsep #2}%
  }

\usepackage{fancyhdr}
\usepackage{lineno}
\usepackage{float}

\usepackage[section]{placeins}
\usepackage[table]{xcolor}

\title{GPU-Accelerated Direct Transcription-Based \\Nonlinear Model Predictive Control}
\author{Evelyn Gondosiswanto$^1$ and Joshua L. Pulsipher$^1$\thanks{Corresponding Author: pulsipher@uwaterloo.ca}\\
    {\small $^1$Department of Chemical Engineering}\\
	{\small \;University of Waterloo, 200 University Ave W, Waterloo, ON N2L 3G1, Canada}}
\date{}

\begin{document}

\maketitle

\begin{abstract}
In this paper, we present a GPU-accelerated framework for nonlinear model predictive control (NMPC) based on direct transcription and second-order interior-point methods. Many real-world systems exhibit nonlinear dynamics that cannot be accurately captured by linear models, motivating the use of NMPC. However, NMPC requires the repeated real-time solution of optimal control problems (OCP), which become computationally demanding large-scale nonlinear programs (NLPs) after transcription. Although GPU acceleration has emerged as a promising approach for nonlinear optimization, existing GPU-based NMPC workflows reconstruct structurally identical OCPs at each solve. This introduces substantial overhead even though successive solves differ only through updated system measurements or reference trajectories. To address this limitation, we introduce a parametric interior-point formulation that exploits the fixed structure of transcribed OCPs, enabling reuse of structure-dependent computations (e.g., symbolic factorization in sparse Cholesky) across re-solves. We evaluate the proposed framework on distillation column and 2D heated plate benchmarks against state-of-the-art CPU and GPU configurations. The results show that the framework achieves over an order-of-magnitude speedup in total NMPC run times. These improvements are primarily driven by reduced per-iteration solve times, with GPU execution achieving up to a $94\%$ reduction compared to the baseline. Overall, the results demonstrate the effectiveness of exploiting repeated problem structure in GPU-accelerated NMPC and highlight the potential of the proposed framework to expand the envelope of real-time NMPC applications.
\end{abstract}

\noindent\textbf{Keywords:} Infinite-Dimensional Optimization; Dynamic Optimization; PDE-Constrained Optimization; Model Predictive Control

\section{Introduction}
Model predictive control (MPC) is a widespread strategy for controlling constrained multivariable systems \cite{forbes2015industry}. Originally developed for power plants and petroleum refineries \cite{qin2000nmpc, schwenzer2021mpc}, MPC has since expanded to numerous applications including aerospace \cite{eren2017aerospace, chai2020mpc}, robotics \cite{falanga2018mpc, grandia2019mpc}, automotive systems \cite{yurtsever2020auto, adabag2025toyota, pratama2024auto} and HVAC control \cite{taheri2022hvac, lazic2018hvac, afram2017hvac}. MPC repeatedly solves an optimal control problem (OCP) that predicts system behavior using a mathematical model and computes control inputs that optimize a specified objective (e.g., setpoint tracking) \cite{schwenzer2021mpc}. Between MPC iterations, the OCP retains an identical problem structure and differs only through updated problem data, such as state measurements and setpoint updates. Consequently, the OCP must be solved in real-time to apply control inputs before the system dynamics evolve significantly. Linear MPC (LMPC), which employs a linear model and quadratic objective, is commonly implemented for its computational efficiency \cite{forbes2015industry}. However, many real-world processes exhibit nonlinear dynamics that cannot be adequately captured by linear approximations \cite{qin2000nmpc, allgower2004nmpc}. Nonlinear MPC (NMPC) addresses this by incorporating nonlinear models and objectives for improved accuracy, but at a higher computational cost \cite{allgower2004nmpc}.

Numerical approaches for solving OCPs are generally classified as either indirect and direct methods \cite{benson2006direct}. Indirect methods (e.g., shooting) derive the optimality conditions and solve the resulting boundary value problem, whereas direct methods discretize the OCP itself and solve the resulting nonlinear program (NLP) \cite{benson2006direct}. In this work, we focus on direct transcription approaches that discretize the state and control trajectories while enforcing system dynamics via differential and algebraic constraints. Since OCPs are inherently infinite-dimensional, being defined over continuous-time (and sometimes space or uncertainty as well), discretization produces a sparse, large-scale NLP typically solved using methods such as interior-point \cite{pulsipher2024scalable, wachter2006ipopt}. Solving large-scale NLPs is numerically demanding, posing a key challenge of real-time NMPC implementations \cite{pulsipher2024scalable}.


To overcome the computational burden of NMPC, several strategies have been explored. Decomposition methods partition the OCP into smaller, tractable subproblems \cite{forbes2015industry, marti2013distribute, liu2009distribute} or separate linear subproblems from the nonlinear ones to reduce complexity \cite{zhu2000decomposition}. However, such approaches may not fully capture subsystem interactions when computing the optimal solution and are often difficult to implement for general OCPs. Data-driven methods have also gained traction, with neural networks (NNs) acting as surrogate models when mechanistic models are unavailable or costly to implement \cite{piche2000neural, schweidtmann2021neural}. Neural ordinary differential equations (ODE) \cite{luo2023node, chee2023node} learn complex continuous-time dynamics from process data, but may generalize poorly under operating conditions outside the data scope. Alternatively, hybrid approaches combine mechanistic and data-driven models to improve prediction accuracy while retaining greater interpretability than pure data-driven approaches \cite{bradley2022perspectives, shah2025hybrid}. Another direction is physics-informed neural networks (PINNs), which embed physical laws into the training process to improve generalizability, though constraint satisfaction is not guaranteed \cite{raissi2019pinn, antonelo2024pinn, zhong2025pinn, zheng2023pinn}. Despite these advances, data-driven methods are fundamentally dependent on training data quality and do not necessarily outperform the original mechanistic models in terms of computational cost for NMPC \cite{casas2025pinn}.

More recently, GPU-accelerated optimization has emerged as a promising approach for large-scale NLPs \cite{shin2024opf, pacaud2024gpu, montoison2025ocp, adabag2024gpu, gondosiswanto2025advances}. Transcribed NLPs exhibit a recurrent algebraic structure (e.g., enforcing path constraints at each discretization point), making them well-suited for GPU parallelization \cite{adabag2024gpu}. Several works exploit this structure to solve NLPs more efficiently. Adabag et al. develop \texttt{MPCGPU}, a GPU-based NMPC framework that employs a first-order preconditioned conjugate gradient (PCG) method \cite{adabag2024gpu}. While computationally lightweight, first-order methods often require many iterations to converge. In contrast, Shin et al. present the single instruction, multiple data (SIMD-NLP) abstraction implemented in \texttt{ExaModels.jl}, coupled with the second-order interior-point solver \texttt{MadNLP.jl} and NVIDIA's sparse Cholesky solver \texttt{cuDSS} \cite{shin2024opf}. By leveraging Hessian information, the approach converges in fewer, albeit more expensive, iterations. This tradeoff between iteration cost and iteration count is particularly important for real-time NMPC applications. Building on \cite{shin2024opf}, Montoison et al. apply the SIMD-NLP framework to OCPs through \texttt{OptimalControl.jl} \cite{montoison2025ocp}. Pulsipher and Shin extend the abstraction to infinite-dimensional optimization (InfiniteOpt) problems in \cite{pulsipher2024scalable}, leading to the InfiniteSIMD-NLP abstraction implemented in \texttt{InfiniteExaModels.jl} \cite{gondosiswanto2025advances}. Combined with \texttt{MadNLP.jl}, benchmarks on stochastic, dynamic and partial differential equation (PDE)-constrained optimization problems demonstrate speedups of one to two orders-of-magnitude over traditional CPU-based workflows \cite{gondosiswanto2025advances}. 

While GPU acceleration has significantly improved OCP solution times, existing approaches in the literature have only focused on one-off solves and do not exploit the recurrent structure arising from repeated OCP solves. Consequently, this presents a performance bottleneck arising from repeated problem initialization \cite{pacaud2024gpu, gondosiswanto2025advances}. In principle, structurally-dependent solution steps (e.g., symbolic factorization in sparse Cholesky \cite{pacaud2024gpu}) can be computed once and reused across NMPC iterations. However, this has not been explored or implemented in existing GPU-NMPC frameworks. In this work, we propose a second-order GPU-accelerated NMPC framework that exploits recurrent structure across re-solves through a parametric optimal control formulation. The main contributions are summarized below:
\begin{itemize}

\item{Develop a parametric modeling abstraction that leverages recurrent algebraic structure for efficient parallelized computations when modeling and solving NMPC problems on GPU. Although this work focuses GPU acceleration, the abstraction is general and applies to CPU-based workflows as well.}

\item{Formalize a parametric, GPU-optimized interior-point algorithm for efficiently re-solving transcribed OCPs by exploiting fixed problem structure across NMPC iterations.}

\item{Provide a software implementation of the proposed framework via \texttt{InfiniteOpt.jl} and \texttt{InfiniteExaModels.jl}, where OCP structure is constructed once and reused across re-solves. Both GPU and CPU-based workflows are supported.}

\item{Benchmark different framework configurations against state-of-the-art CPU and GPU approaches to demonstrate how our framework achieves up to a 94\% reduction in solve time for two NMPC case studies.}

\end{itemize}

The rest of this paper is as follows. Section \ref{sec:background} introduces relevant notation and background. Section \ref{sec:framework} details the framework's workflow and its implementation in the \texttt{InfiniteOpt.jl} and \texttt{InfiniteExaModels.jl} packages. The case studies and their performance benchmarks are presented in Section \ref{sec:caseStudies}. Lastly, concluding remarks and future directions are given in Section \ref{sec:conclusion}.

\section{Notation and Background} \label{sec:background}
InfiniteOpt problems (e.g., OCPs) involve decision variables that are defined over continuous domains $\mathcal{D}$ such as time or space \cite{pulsipher2022unifying}. In \cite{pulsipher2022unifying}, Pulsipher et al. present a unifying modeling abstraction for these problems which are expressed using four core modeling objects: infinite parameters $d \in \mathcal{D} \subseteq \mathbb{R}^{n_d}$, infinite variables $y : \mathcal{D} \mapsto \mathcal{Y} \subseteq \mathbb{R}^{n_i}$, measure operators $\mathcal{M} : \mathscr{Y} \mapsto \mathbb{R}$, and differential operators $D : \mathscr{Y} \mapsto \mathscr{D}$ where $\mathscr{Y}$ is the function space describing $y(d)$. Together, these produce a general formulation for InfiniteOpt problems:

\begin{equation} \label{eq:infiniteOpt_unifying}
\begin{aligned}
    && \min_{y, y_f} &&& \mathcal{M}_d f(Dy, y(d), y_f, d)\\
    && \text{s.t.} &&& g(Dy, y(d), y_f, d) \leq 0, && d \in \mathcal{D}\\
    &&&&& h(Dy, y(d), y_f, d) = 0, && d \in \mathcal{D}
\end{aligned}
\end{equation}

\noindent where $f(\cdot)$ is the objective function, $g(\cdot) \leq 0$ capture inequality constraints (e.g., path constraints), $h(\cdot) = 0$ encompass equality constraints such as differential-algebraic equation (DAE) constraints, and $y_f$ are finite-dimensional variables. This abstraction captures InfiniteOpt problems that arise from a wide range of applications, including stochastic, dynamic, and PDE-constrained optimization \cite{pulsipher2022unifying}. For example, a dynamic optimization problem is formulated by using a time domain $\mathcal{D}_t = [t_0, t_f]$.

To solve with nonlinear solvers like \texttt{Ipopt}, InfiniteOpt problems are transcribed into a compatible finite-dimensional representation via direct transcription. By discretizing the continuous domain(s) $\mathcal{D}$ into a finite set of discretization points $\hat{\mathcal{D}} := \{\hat{d}:r \in \mathcal{R}\} \subset \mathcal{D}$, Problem \eqref{eq:infiniteOpt_unifying} turns into the nonlinear program (NLP) shown below:

\begin{equation} \label{eq:infiniteOpt_discrete}
\begin{aligned}
    && \min &&& \sum_{r \in \mathcal{R}} \phi_r f(\hat{Dy}_r, \hat{y}_r, y_f, \hat{d}_r)\\
    && \text{s.t.} &&& g(\hat{Dy_r}, \hat{y}_r, y_f, \hat{d}_r) \leq 0, && r \in \mathcal{R}\\
    &&&&& h(\hat{Dy_r}, \hat{y}_r, y_f, \hat{d}_r) = 0, && r \in \mathcal{R}\\
    &&&&& q(\hat{Dy}, \hat{y}, \hat{d}) = 0\\
\end{aligned}
\end{equation}

\noindent where $\mathcal{M}_d$ is approximated via a weighted sum using weights $\phi_r$ from an appropriate numerical scheme (e.g., trapezoidal rule). While $h(\cdot)$ encapsulates equality constraints like boundary conditions, $q(\cdot)$ encodes numerical approximations of the differential variables $\hat{Dy}$ via methods such as finite difference or orthogonal collocation over finite elements \cite{pulsipher2022unifying}. The functions $f(\cdot), g(\cdot), h(\cdot), q(\cdot)$ are repeated over $\hat{\mathcal{D}}$ which is often a large set, exhibiting a highly recurrent algebraic structure \cite{pulsipher2024scalable}. Such large-scale NLPs are computationally demanding to solve in real-time, especially with the presence of PDEs.

Problem \eqref{eq:infiniteOpt_discrete} is modeled using algebraic modeling languages (AMLs) like \texttt{JuMP}, \texttt{Pyomo} or \texttt{AMPL}, which often compute derivatives for NLP solvers via automatic differentiation (AD) \cite{dunning2017jump, nicholson2018pyomo, gay2015ampl}. While computationally efficient, AD applies the chain rule over expression trees for every instance of the constraint and objective terms, leading to redundant evaluations across all discretization points \cite{pulsipher2024scalable, margossian2019ad}. To exploit recurrent structure, we can instead transcribe Problem \eqref{eq:infiniteOpt_unifying} using the InfiniteSIMD-NLP abstraction \cite{gondosiswanto2025advances}:

\begin{equation} \label{eq:infinitesimd_form}
\begin{aligned}
    && \min &&& \sum_{r \in \mathcal{R}} \tilde{f}(\hat{Dy}, \hat{y}, y_f; r, \phi_r, \hat{d}_r)\\
    && \text{s.t.} &&& \tilde{g}_{\upsilon}(\hat{Dy}, \hat{y}, \hat{s}, y_f; r, \hat{d}_r) + \sum_{\kappa \in \mathcal{K}_{\upsilon}} \sum_{\tau\in T_{\kappa}} \tilde{g}'_{\kappa} (\hat{Dy}, \hat{y}, y_f; r, \hat{d}_r, c_{\tau \kappa}) = 0, && \upsilon \in \Upsilon, r \in \mathcal{R}\\
    &&&&& \tilde{h}_{\theta}(\hat{Dy}, \hat{y}, \hat{s}, y_f; r, \hat{d}_r) + \sum_{\kappa \in \mathcal{K}_{\theta}} \sum_{\tau\in T_{\kappa}} \tilde{h}'_{\kappa} (\hat{Dy}, \hat{y}, y_f; r, \hat{d}_r, \epsilon_{\tau \kappa}) = 0, && \theta \in \Theta, r \in \mathcal{R}\\
    &&&&& \tilde{q}(\hat{Dy}, \hat{y}; \hat{d}, a_{\iota}) + \sum_{\rho \in \mathcal{P}} \tilde{q}'(\hat{Dy}, \hat{y}; \hat{d}, b_{\rho}) = 0, &&  \iota \in \mathcal{I}
\end{aligned}
\end{equation}

\noindent where the functions $\tilde{f}(\cdot), \tilde{g}(\cdot), \tilde{g}'(\cdot), \tilde{h}(\cdot), \tilde{h}'(\cdot), \tilde{q}(\cdot)$ and $\hat{q}'(\cdot)$ represent distinct algebraic patterns evaluated across discretization points indexed by $r$, with fixed structure that varies only through data inputs $c, \epsilon, a, b$ across instances indexed by $\kappa, \iota, \rho$. Building on the SIMD-NLP abstraction, symbolic AD derives symbolic function and derivative expressions once for each unique algebraic pattern \cite{pacaud2024gpu}. These expressions are then compiled into reusable computational kernels $\mathcal{K}_{SIMD}$ that are evaluated with different data inputs across all discretization points to enable efficient AD. Although this approach can accelerate AD on CPUs, GPU architectures are inherently tailored for SIMD computing tasks, making InfiniteOpt problems transcribed in the form of Eqn. \eqref{eq:infinitesimd_form} well-suited for GPU parallelization \cite{pulsipher2024scalable, gondosiswanto2025advances}.

The unifying abstraction in \eqref{eq:infiniteOpt_unifying} is implemented in \texttt{InfiniteOpt.jl}, which builds on \texttt{JuMP.jl} \cite{lubin2023jump} to support infinite-dimensional modeling objects. InfiniteOpt problems are created/stored as \texttt{InfiniteModel}s and automatically undergo direct transcription via a transformation backend, which produces transcribed \texttt{JuMP} models by default. \texttt{InfiniteOpt.jl} also supports the InfiniteSIMD-NLP abstraction via \texttt{InfiniteExaModels.jl}, where InfiniteOpt problems are transcribed with an \texttt{ExaTranscriptionBackend} into a SIMD-based \texttt{ExaModel} \cite{pulsipher2024scalable, gondosiswanto2025advances}. When paired with \texttt{MadNLP.jl} and \texttt{cuDSS}, this workflow achieves speedups of at least one to two orders-of-magnitude on dynamic, PDE and stochastic InfiniteOpt benchmarks compared to state-of-the-art CPU approaches, exhibiting near-linear scaling with problem size \cite{gondosiswanto2025advances}. However, these results are based on single NLP solves and do not consider the repeated solve paradigm used in NMPC where reinitialization costs can become a bottleneck for real-time solution \cite{gondosiswanto2025advances}. Section \ref{sec:framework} details a proposed framework to amortize these costs across NMPC iterations.

\section{GPU-NMPC Framework} \label{sec:framework}
In this section, we detail the proposed GPU-NMPC framework and its implementation in \texttt{InfiniteOpt.jl} and \texttt{InfiniteExaModels.jl}. The framework is based on an NMPC loop that repeatedly solves the OCP for optimal control inputs, applies the inputs to the system, and updates the state measurements between iterations, as presented in Algorithm \ref{algo-nmpc}. In each NMPC iteration, the OCP is solved using a parametric interior-point method detailed in Section \ref{sec:parametric-ipopt} and summarized in Algorithm \ref{algo-interior}. 

\subsection{Problem Definition and Transcription}
Consider an NMPC problem defined over the continuous-time domain $\mathcal{T} = [0, t_{f}]$ with $N_{mpc}$ control steps. At each control step $t_k$ for $k \in \{1, 2, ..., N_{mpc}\}$, we solve an OCP $\mathcal{P}_k(p_k(t))$ with input parameters $p_k(t)$ over the prediction horizon $\mathcal{T}_p = [0, t_p]$ and control horizon $\mathcal{T}_c = [0, t_c]$, where $t_c \leq t_p$:

\begin{equation} \label{eq:infiniteOCP-continuous}
\begin{aligned}
\mathcal{P}_{k}(p_{k}(t)) = && \min_{y} &&& \mathcal{M}_{t, d} f(Dy, y(t, d), p_{k}(t),  t, d)\\
&& s.t. &&& g(Dy, y(t, d), p_{k}(t), t, d) \leq 0, && t \in \mathcal{T}_p, d \in \mathcal{D}\\
&&&&&  h(Dy, y(t, d), p_{k}(t), t, d) = 0, && t \in \mathcal{T}_p, d \in \mathcal{D}\\
&&&&& y_s(0, d) = \bar{y_s}(p_{k}), && d \in \mathcal{D}\\
&&&&& y_c(t, d) = y_c(t_c, d), && t \in \mathcal{T}_p \setminus \mathcal{T}_c, d \in \mathcal{D}\\
\end{aligned}
\end{equation}

\noindent where $t \in \mathcal{T}_p$ denotes the time domain and $d \in \mathcal{D}$ represents additional domains such as space ($x \in \mathcal{D}_x$) or uncertainty ($\xi \in \mathcal{D_\xi}$). The function $y(t, d) = (y_s(t,d), y_c(t, d), y_f)$ consists of the state variables $y_s$, control inputs $y_c$ and finite decision variables $y_f$. If $\mathcal{D}$ is empty, then $y_s$ and $y_c$ depend only on time. For example, in a heating process, $y_s(t, x)$ can denote the temperature distribution, $y_c(t, x)$ can describe the distributed heating control policy, and $y_f$ can correspond to settings that are space-time invariant. The measure operator $\mathcal{M}_{t, d}$ summarizes the objective function $f$ over the continuous domains (e.g., a Bolza objective), while $g(\cdot)$ and $h(\cdot)$ denote (nonlinear) DAEs, path constraints, and boundary conditions. Initial conditions are specified via $\bar{y_s}: \mathbb{R}^{n_p} \rightarrow \mathbb{R}^{n_s}$, which extracts the relevant initial state values from $p_{k}$. The control horizon restriction is enforced by holding $y_c(t, d)$ constant for $t \geq t_c$. Problem \eqref{eq:infiniteOCP-continuous} is parametrized by the input vector $p_{k}(t)$, which describes the OCP parameters at time $t_k$. Such parameters may either vary over the infinite domain (e.g., a time-varying setpoint) or remain constant such as state measurements (i.e., initial conditions). For each NMPC iteration $k$, $\mathcal{P}_k$ is updated through $p_k(t)$ while retaining the same objective and constraint functions.

To solve Problem \eqref{eq:infiniteOCP-continuous}, we apply direct transcription by discretizing the prediction horizon as $\hat{\mathcal{T}}_p := \{\hat{t}_u : u \in \mathcal{U} \} \subset \mathcal{T}_p$ and the control horizon as  $\hat{\mathcal{T}}_c := \{\hat{t}_u : u \in \mathcal{U}_c\} \subseteq \hat{\mathcal{T}}_p$ where $\mathcal{U}_c = \{u: \hat{t}_u \in \mathcal{T}_c, u \in \mathcal{U}\}$. Additional domains are treated in the same manner to get $\hat{\mathcal{D}} := \{\hat{d} : r \in \mathcal{\mathcal{R}}\} \subset \mathcal{D}$. Problem \eqref{eq:infiniteOCP-continuous} is thus transformed into the following NLP (i.e., the transcribed OCP $\hat{\mathcal{P}}_k(\hat{p}_{k})$):

\begin{equation} \label{eq:infiniteOCP-discrete}
\begin{aligned}
\hat{\mathcal{P}}_{k}(\hat{p}_{k}) = && \min &&& \sum_{u \in \mathcal{U}} \sum_{r \in \mathcal{R}} \phi_{ur} f(\hat{y}_{ur}, \hat{p}_{ku}, \hat{t}_u, \hat{d}_r)\\
&& s.t. &&& g(\hat{y}_{ur}, \hat{p}_{ku}, \hat{t}_u, \hat{d}_r) \leq 0, \quad u \in \mathcal{U}, r \in \mathcal{R}\\
&&&&& h(\hat{y}_{ur}, \hat{p}_{ku}, \hat{t}_u, \hat{d}_r) = 0, \quad u \in \mathcal{U}, r \in \mathcal{R}\\
&&&&& q(\hat{y}_{ur}, \hat{p}_{ku}, \hat{t}_u, \hat{d}_r) = 0, \quad u \in \mathcal{U}, r \in \mathcal{R}\\
&&&&& y_{s, 0r} = \bar{y_s}(\hat{p}_{k0}), \quad r \in \mathcal{R}\\
&&&&& y_{c, ur} = y_{c, u_cr}, \quad u \in \mathcal{U} \setminus \mathcal{U}_c, r \in \mathcal{R}\\
\end{aligned}
\end{equation}

\noindent where $\hat{y}_{ur} = (\hat{D}y_{s, ur}, \hat{y}_{s, ur}, \hat{y}_{c, ur}, y_f)$ denotes the discretized variables evaluated at the transcription point $(\hat{t}_u, \hat{d}_r)$ and $u_c$ is the index that corresponds to $\hat{t}_{u = u_c} = t_c$. The vector $\hat{p}_k$ collects the parameter values over all discretized time points, while $\hat{p}_{ku}$ denotes the parameter values corresponding to $\hat{t}_u$ specifically. The measure operator $\mathcal{M}_{t, d}$ from Problem \eqref{eq:infiniteOCP-continuous} is approximated with weights $\phi_{ur}$, and $q(\cdot)$ are the auxiliary derivative equations corresponding to $\hat{Dy}_s$ \cite{pulsipher2022unifying}. As $\hat{\mathcal{P}}_k(\hat{p}_k)$ typically contains few unique algebraic constraints over the discretized domain(s), Problem \eqref{eq:infiniteOCP-discrete} often exhibits a highly sparse and repetitive structure that can be equivalently represented using the InfiniteSIMD-NLP abstraction \eqref{eq:infinitesimd_form} for GPU execution, as described in Section \ref{sec:background}. The repeated inequality structures in $g(\cdot)$ are expressed through the SIMD-compatible functions $\tilde{g}(\cdot)$ and $\tilde{g}'(\cdot)$. Similarly, the equality constraints $h(\cdot)$, the initial conditions, and the control constraints are represented through $\tilde{h}(\cdot)$ and $\tilde{h}'(\cdot)$. The differential constraints $q(\cdot)$ are likewise represented through $\tilde{q}(\cdot)$ and $\tilde{q}'(\cdot)$. For compactness in notation, we rewrite Problem \eqref{eq:infiniteOCP-discrete} as:
\begin{equation} \label{eq:infiniteOCP-general}
\begin{aligned}
\hat{\mathcal{P}}(p) = && \min_{y \in \mathbb{R}^n, s \in \mathbb{R}^{m_g}} &&& \hat{f}(\hat{y}; p)\\
&& s.t. &&& \hat{h}(\hat{y}; p) = 0\\
&&&&& \hat{g}(\hat{y}; p) + s = 0\\
&&&&& s \geq 0\\
\end{aligned}
\end{equation}

\noindent where $\hat{f}: \mathbb{R}^n \rightarrow \mathbb{R}$ aggregates the weighted discretized objective terms. We introduce $\hat{h}(\cdot): \mathbb{R}^n \rightarrow \mathbb{R}^{m_h}$ to group all the equality constraints, including $h(\cdot)$, derivative approximations $q(\cdot)$, initial conditions and the control-horizon constraints.  More precisely, $\hat{h}_{ur}(\cdot)$ represents the individual equality constraints that correspond to the point $(\hat{t}_u, \hat{d}_r) \in \hat{\mathcal{T}}_p \bigtimes \hat{\mathcal{D}}$. The full vector $\hat{h}(\cdot)$ is then obtained by stacking all the equality constraints across $\hat{\mathcal{T}}_p \bigtimes \hat{\mathcal{D}}$. Similarily, $\hat{g}(\cdot): \mathbb{R}^n \rightarrow \mathbb{R}^{m_g}$ is formed by stacking the discretized inequality constraints $\hat{g}_{ur}(\cdot)$, which are reformulated into equalities using slack variables $s$. We also denote the Lagrange multipliers $\lambda \in \mathbb{R}^{m_h}$, $\omega \in \mathbb{R}^{m_g}$ for the equality and inequality constraints, respectively, and the multipliers for the slack variables as $\beta \in \mathbb{R}^{m_g}$. Using this notation, the Lagrangian $\mathcal{L}$ for Problem \eqref{eq:infiniteOCP-general} is given by:

\begin{equation} \label{eq:infiniteOCP-lagrangian}
\mathcal{L}(\hat{y}, s, \lambda, \omega, \beta; p) = \hat{f}(\hat{y}; p) + \lambda^T \hat{h}(\hat{y}; p) + \omega^T(\hat{g}(\hat{y}; p) + s) - \beta^Ts.\\
\end{equation}

The corresponding first-order Karush-Kuhn-Tucker (KKT) optimality conditions then follow as:
\begin{equation} \label{eq:infiniteOCP-kkt}
\begin{gathered}
\nabla_y\mathcal{L}(\hat{y}, s, \omega, \lambda, \beta; p) = 0 \\
\omega - \beta = 0 \\
\hat{h}(\hat{y}; p) = 0\\
\hat{g}(\hat{y}; p) + s = 0 \\
0\leq s \perp \beta \geq 0
\end{gathered}
\end{equation}

\noindent where $\perp$ is used to denote a complementarity condition. The primal-dual variable $z := (\hat{y}, s, \omega, \lambda, \beta)$ is a solution to Problem \eqref{eq:infiniteOCP-general} if it satisfies \eqref{eq:infiniteOCP-kkt}. Thus, solving \eqref{eq:infiniteOCP-kkt} is equivalent to solving \eqref{eq:infiniteOCP-general}. As $\hat{\mathcal{P}}(p)$ is repeatedly solved within the NMPC loop, it maintains a fixed problem structure as $p$ is updated at each $t_k$. This motivates solution strategies with efficient parameter updates on GPU, as shown in Section \ref{sec:parametric-ipopt}.
 
\subsection{Parametric Interior-Point Method on GPU for NMPC} \label{sec:parametric-ipopt}
This section presents a GPU-compatible parametric interior-point formulation for efficient re-solves of Problem \eqref{eq:infiniteOCP-general}. By exploiting the fixed structure of the parametric KKT conditions in \eqref{eq:infiniteOCP-kkt}, the proposed approach reuses structure-dependent computations across NMPC iterations. While approaches have been proposed in the literature for individual NLP solves \cite{pacaud2024gpu}, our approach extends these to explicitly account for parametric updates to a fixed transcribed OCP structure, enabling efficient re-solves as shown in Algorithm \ref{algo-interior}.

\subsubsection{KKT System Setup}
 The non-smooth KKT conditions in \eqref{eq:infiniteOCP-kkt} are reformulated using a homotopy method, yielding the smooth KKT system $K_\mu(z; p)$:
 
\begin{equation} \label{eq:infiniteOCP-smoothkkt}
K_\mu(z; p) = 
\begin{bmatrix}
\nabla_y\mathcal{L}(\hat{y}, s, \omega, \lambda, \beta; p)\\
\hat{h}(\hat{y}; p)\\
\hat{g}(\hat{y}; p) + s\\
S\beta - \mu e_{g}\\
\end{bmatrix}
\end{equation}

\noindent where $e_{g} \in \mathbb{R}^{m_g}$ is a vector of ones, $\mu > 0$ is the barrier parameter, and $S$ = diag($s$) is a diagonal matrix. The system $K_\mu(z;p) = 0$ is solved using an interior-point method over an "outer loop" sequence of decreasing barrier parameters $\mu \rightarrow 0$. At each outer loop iteration indexed by $j$, the corresponding smooth KKT system is solved to obtain the barrier solution $z_{\mu_j}^*$. As the barrier parameter decreases, the sequence of barrier solutions converges to the optimal solution $z_{\mu}^*$ of $\hat{\mathcal{P}}_k(\hat{p}_k)$. Note that when $\mu = 0$, $K_\mu(z;p)$ becomes the KKT conditions for the original problem.

\subsubsection{Augmented Primal-Dual KKT System}
For a fixed $\mu_j$, an "inner loop" solves Eqn. \eqref{eq:infiniteOCP-smoothkkt} using Newton's method \cite{wachter2006ipopt}. At each inner loop iteration indexed by $i$, $K_\mu(z; p)$ is linearized at the point $z_{\mu_ji} = (\hat{y}_i, s_i, \lambda_i, \omega_i)$ to obtain the augmented KKT system $K_{aug}(z_{\mu_ji}; p)$:

\begin{equation} \label{eq:infiniteOCP-augmented}
\overbracket{
\begin{bmatrix}
W_i(z_{\mu_ji}; p) & 0 & H_i(\hat{y}_i; p)^T & G_i(\hat{y}_i; p)^T\\
0 & Q_s & 0 & I\\
H_i(\hat{y}_i; p) & 0 & 0 & 0\\
G_i(\hat{y}_i; p) & I & 0 & 0\\
\end{bmatrix}
}^{K_{aug}(z_{\mu_ji}; p)}
\begin{bmatrix}
\delta_y\\
\delta_s\\
\delta_\lambda\\
\delta_\omega\\
\end{bmatrix} = - \vec{r}(z_{\mu_ji}; p)
\end{equation}

\noindent where $Y$ = diag($\hat{y}$), $\mathcal{B}$ = diag($\beta$), $Q_s = S_i^{-1}\mathcal{B}_i$, $W_i(z_{\mu_ji}; p) = \nabla_{yy}^2\mathcal{L}(z_{\mu_ji}; p)$, $H_i(\hat{y}_i; p) = \nabla_y \hat{h}(\hat{y}_i; p)$, $G_i(\hat{y}_i; p) = \nabla_y \hat{g}(\hat{y}_i; p)$. The gradient terms $W_i(\cdot), H_i(\cdot)$ and $G_i(\cdot)$ are efficiently evaluated in parallel using $\mathcal{K}_{SIMD}$. We denote the right hand side of the system as $\vec{r} = [r_1, r_2, r_3, r_4]$ which is defined as:

\begin{equation} \label{eq:infiniteOCP-augmentedRHS}
\vec{r}(z_{\mu_ji}; p) = 
\begin{bmatrix}
r_1(z_{\mu_ji}; p)\\
r_2(z_{\mu_ji})\\
r_3(z_{\mu_ji}; p)\\
r_4(z_{\mu_ji}; p)\\
\end{bmatrix} =
\begin{bmatrix}
\nabla \hat{f}(\hat{y}_i; p) + \nabla \hat{h}(\hat{y}_i; p)^T\lambda_i + \nabla \hat{g}(\hat{y}_i; p)^T\omega_i - \mu Y^{-1}e\\
\omega_i - \mu S^{-1}e\\
\hat{h}(\hat{y}_i; p)\\
\hat{g}(\hat{y}_i; p) + s_i\\
\end{bmatrix}.
\end{equation}

\noindent Note that the function and gradient evaluations are also efficiently computed using the $\mathcal{K}_{SIMD}$ set up through the InfiniteSIMD-NLP representation of Problem \eqref{eq:infiniteOCP-general}.
Equation \eqref{eq:infiniteOCP-augmented} is solved to obtain the search directions $\delta_i = (\delta_y, \delta_s, \delta_\lambda, \delta_\omega)$ for Newton's method. Since the system incorporates second-order information through the Hessian $W_i(z_{\mu_ji}; p)$, interior-point methods are classified as second-order methods. For the solution $\delta_i$ to exist, $K_{aug}(z_{\mu_ji}; p)$ must be invertible. This can be checked by looking at the inertia:

\begin{equation} \label{infiniteOCP-inertia}
\text{inertia}(K_{aug}(z_{\mu_ji}; p)) = (n_+, n_0, n_-)
\end{equation}

\noindent where $n_+, n_0, n_-$ denote the number of positive, null and negative eigenvalues, respectively. If inertia$(K_{aug}(z_{\mu_ji}; p)) = (n + m_g, 0, m_g + m_h)$, then $K_{aug}(z_{\mu_ji}; p)$ is invertible and the search directions can be computed \cite{wachter2006ipopt, pacaud2024gpu}. Otherwise, prime and dual regularization parameters $(\eta_y, \eta_c) > 0$ are computed and introduced into $K_{aug}(z_{\mu_ji}; p)$ to make it invertible:

\begin{equation} \label{eq:infiniteOCP-regularize}
K_{aug}(z_{\mu_ji}; p) = \begin{bmatrix}
W_i(z_{\mu_ji}; p) + \eta_yI & 0 & H_i(\hat{y}_i; p)^T & G_i(\hat{y}_i; p)^T\\
0 & Q_s + \eta_yI & 0 & I\\
H_i(\hat{y}_i; p) & 0 & -\eta_cI & 0\\
G_i(\hat{y}_i; p) & I & 0 & -\eta_cI\\
\end{bmatrix}.
\end{equation}

Nonlinear solvers commonly solve Eqn. \eqref{eq:infiniteOCP-augmented} using a sparse $LDL^T$ factorization routine, which decomposes the matrix into a lower triangular matrix $L$ and a block diagonal matrix $D$ \cite{pacaud2025kkt}. However, algorithms that implement $LDL^T$ factorization use numerical pivoting which is not readily amendable to parallelization, making it not well-suited for GPU architectures \cite{pacaud2024gpu}. Instead, we further treat the KKT system with a condensation step, setting $K_c(z_{\mu_{j}i};p) := W_i + \eta_yI + G_i^TQ_sG_i$ and eliminating the block rows associated with $\delta_s$ and $\delta_\omega$:
\begin{equation} \label{eq:infiniteOCP-condensed}
\begin{bmatrix}
K_c(z_{\mu_{j}i};p) & H_i(\hat{y}_i; p)^T\\
H_i(\hat{y}_i; p) & 0\\
\end{bmatrix}
\begin{bmatrix}
\delta_y\\
\delta_\lambda\\
\end{bmatrix} = -
\begin{bmatrix}
r_1 + G_i(\hat{y}_i; p)^T (Q_sr_4 - r_2)\\
r_3\\
\end{bmatrix}.
\end{equation}

By Sylvester's law of inertia, the inertias of $K_{aug}$ and $K_c$ satisfy the relation \cite{pacaud2025kkt}:

\begin{equation} \label{eq:infiniteOCP-inertia}
\text{inertia}(K_{aug}) = (n + m_g, 0, m_g + m_h) \Leftrightarrow \text{inertia}(K_c) = (n, 0, m_h).
\end{equation}

\noindent Thus, inertia($K_{c}$) can help to indirectly check inertia($K_{aug}$) and determine suitable values for $\eta_y$ and $\eta_c$. Equation \eqref{eq:infiniteOCP-condensed} is solved using a sparse Cholesky factorization which is more amenable to GPU parallelization \cite{pacaud2024gpu}. However, sparse Cholesky requires a positive-definite (PD) matrix and $K_{c}(z_{\mu_{j}i}; p)$ is often not PD. To address this, the system can be reformulated using either Lifted-KKT or HyKKT \cite{pacaud2024gpu}. Lifted-KKT introduces auxiliary variables to lift the system into a higher-dimensional space \cite{shin2024opf}, whereas HyKKT is a hybrid direct-iterative approach based on an augmented Lagrangian formulation \cite{regev2023hykkt, golub2003hykkt}. For concision in presentation, we limit the scope of this work to Lifted-KKT; however, the proposed framework and its implementation (described in Section \ref{sec:software}) also support HyKKT.

By relaxing the equality constraints associated with $H_i(\hat{y}_i; p)$ and selecting an appropriate value for $\eta_y$, the condensed system $K_c(z_{\mu_{j}i}; p)$ can be reformulated into an $n \times n$ PD system. Specifically, the equality constraints $\hat{h}(\cdot)$ in Problem \eqref{eq:infiniteOCP-general} are relaxed using the parameter $\gamma > 0$:

\begin{equation} \label{eq:liftedkkt}
\begin{aligned}
& \min && \hat{f}(\hat{y}; p)\\
& s.t. && -\gamma \leq \hat{h}(\hat{y}; p) \leq \gamma\\
&&& \hat{g}(\hat{y}; p) \leq 0.
\end{aligned}
\end{equation}

\noindent The relaxed equality constraints are incorporated into $\hat{g}(\cdot)$, yielding the lifted inequality constraint function $g^{l}(\cdot): \mathbb{R}^n \rightarrow \mathbb{R}^{m_{g}^{l}}$, where $m_{g}^{l} = m_g + 2m_h$ with corresponding slack variables $s^{l}$. The resulting problem is solved using the same procedure outlined in Equations \eqref{eq:infiniteOCP-lagrangian} to \eqref{eq:infiniteOCP-condensed}. Quantities associated with the Lifted-KKT reformulation are denoted by the superscript $(\cdot)^{l}$ (e.g., $K_{\mu}^{l}, K_{aug}^{l}, G_i^{l},$ etc...). The condensed Lifted-KKT system is therefore given by:

\begin{equation} \label{eq:liftedkkt-pd}
K_c^{l}(z_{\mu_{j}i}; p)\delta_y^{l} = -r_1^{l} -G_i^{l}(\hat{y}_i; p)^T(Q_s^{l}r_4^{l} - r_2^{l})\\
\end{equation}

\noindent where $K_c^l(z_{\mu_{j}i}; p) := W_i^l + \eta_yI + (G_i^l)^T Q_s^l G_i^l$ . The value of $\eta_y$ is determined using an inertia correction method. Based on inertia($K_c^l(z_{\mu_{j}i}; p)$) and the relation in Equation \eqref{eq:infiniteOCP-inertia}, $\eta_y$ is increased as needed to ensure $K_c^l(z_{\mu_{j}i}; p)$ is PD such that it can be factorized using sparse Cholesky \cite{pacaud2024gpu}.

\subsubsection{Sparse Cholesky Factorization} \label{sec:cholesky}
In this context, the application of Cholesky factorization yields $K_c^l(z_{\mu_ji}; p) = LL^T$, where the lower triangular matrix $L$ is known as the Cholesky factor. Computing $L$ consists of two main phases: symbolic and numerical factorization \cite{davis2016sparse}. Symbolic factorization aims to determine the sparsity structure of $L$, denoted as $\mathcal{S}_L$, together with a permutation matrix $P$:

\begin{equation} \label{eq:cholesky-symCholesky}
P, \mathcal{S}_L = \text{SymCholesky}(K_c^l(z_{\mu_ji}; p))
\end{equation}

\noindent where the sparsity structure $\mathcal{S}_L$ is the ordered set of row-column index pairs $(i, j)$ corresponding to the non-zero entries $L_{ij}$. It serves as a structural blueprint for the in-place computation of the numerical entries during numerical factorization.

While calculating the numerical values, updates from previously computed columns may introduce non-zero entries in $L$ at locations corresponding to zero entries in $K_c^l(z_{\mu_ji}; p)$. These are referred to as fill-in entries, whose locations are determined during symbolic factorization. Consequently, $L$ may become denser than the lower triangular portion of $K_c^l(z_{\mu_ji}; p)$, increasing both memory requirements and factorization cost \cite{davis2016sparse}. To mitigate fill-in, $P$ is applied to $K_c^l(z_{\mu_ji}; p)$, resulting in the reordered condensed KKT system $K_P^l(z_{\mu_ji}; p)$:

\begin{equation} \label{eq:cholesky-permute}
K_P^l(z_{\mu_ji}; p) = PK_c^l(z_{\mu_ji}; p)P^T = L_PL_P^T
\end{equation}

\noindent where multiplying by $P$ on both sides produces a symmetric reordering \cite{scott2024cholesky}. Finding an ordering with minimal fill-in is computationally intractable in general, so heuristic methods like minimum degree and nested dissection are commonly used to find an acceptable $P$ such that $L_P$ is sparser than $L$. \cite{davis2016sparse, lin2005cholesky}.

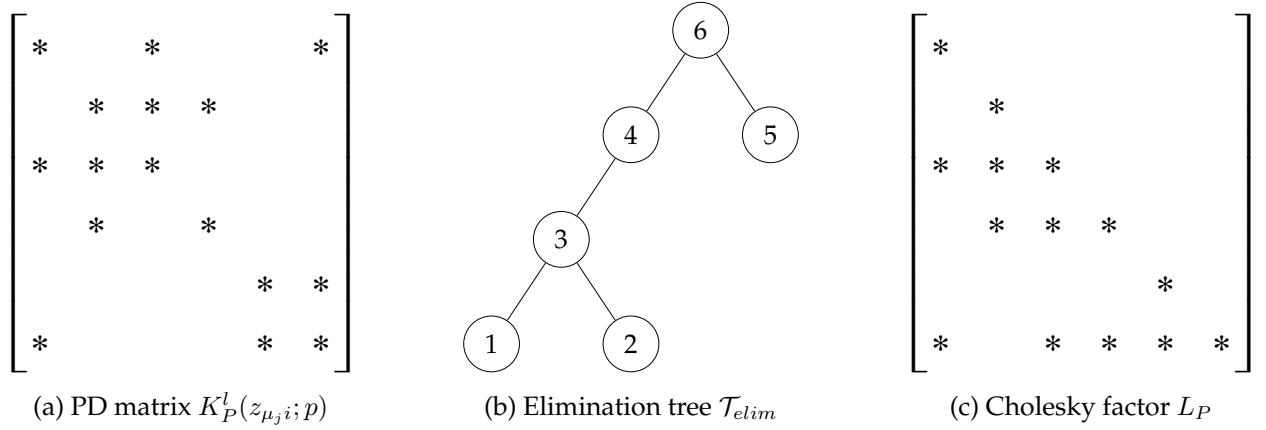
\begin{figure}[!h]
     \centering
     \begin{subfigure}[b]{0.3\textwidth}
         \centering
         \resizebox{!}{2.5cm}{$
         \begin{bmatrix}
            * &  & * &  &  & *\\
            & * & * & * & & \\
            * & * & * & & \\
            & * & & * & \\
            & & & & * & * \\
            * & & & & * & * \\
        \end{bmatrix}
        $}
         \caption{PD matrix $K_P^l(z_{\mu_ji}; p)$}
         \label{fig:cholesky-exmatrix}
     \end{subfigure}
     \hfill
     \begin{subfigure}[b]{0.3\textwidth}
        \centering
        \resizebox{!}{4.9cm}{
         \begin{tikzpicture}[
            node/.style={circle, draw, minimum size=8mm},
            level/.style={dashed, draw, rounded corners, inner sep=6pt}
         ]
        
         \node[node] (6) at (1, 4.5) {6};
        
         \node[node] (4) at (0, 3) {4};
         \node[node] (5) at (2, 3) {5};
         
         \node[node] (3) at (-1, 1.5) {3};
        
         \node[node] (1) at (-2, 0) {1};
         \node[node] (2) at (0, 0) {2};
        
         \draw[-] (6) -- (4);
         \draw[-] (6) -- (5);
         \draw[-] (4) -- (3);
         \draw[-] (3) -- (1);
         \draw[-] (3) -- (2);
        
         \end{tikzpicture}
         }
         \caption{Elimination tree $\mathcal{T}_{elim}$}
         \label{fig:cholesky-etree}
     \end{subfigure}
     \hfill
     \begin{subfigure}[b]{0.3\textwidth}
         \centering
         \resizebox{!}{2.5cm}{$
         \begin{bmatrix}
            * &  &  &  &  &\\
            & * &  & & & \\
            * & * & * & & \\
            & * & * & * & \\
            & & & & * &  \\
            * & & * & * & * & * \\
         \end{bmatrix}
         $}
         \caption{Cholesky factor $L_{P}$}
         \label{fig:cholesky-exfactor}
     \end{subfigure}
    \caption{An example of $K_P^l(z_{\mu_ji}; p)$, its corresponding elimination tree $\mathcal{T}_{elim}$, and Cholesky factor $L_{P}$. Non-zero entries are denoted as * while the column indices define the nodes of $\mathcal{T}_{elim}$.}
    \label{fig:cholesky-symExample}
\end{figure}

After reordering, an elimination tree $\mathcal{T}_{elim}$ is derived from $K_P^l(z_{\mu_ji}; p)$ to capture column dependencies that establish the sparsity pattern of $L_P$. For this discussion, consider an example of $K_P^l(z_{\mu_ji}; p)$ shown in Figure \ref{fig:cholesky-exmatrix} with non-zero entries denoted as *. For each column $j$ in $K_P^l(z_{\mu_ji}; p)$, the row indices of the non-zeros below the diagonal correspond to later columns influenced by column $j$ during factorization. Such dependencies are compactly represented by the elimination tree $\mathcal{T}_{elim}$ in Figure \ref{fig:cholesky-etree}, where each node corresponds to a column of $K_P^l(z_{\mu_ji}; p)$ and the edges denote parent-child dependencies between them. For example, columns 1 and 2 are children of column 3, indicating that both columns structurally influence column 3. The elimination tree also encodes a column-processing order, since child columns must be processed before their parents to determine the corresponding sparsity structures.

The sparsity structure of $L_{P}$ is computed column by column using the dependencies encoded in $\mathcal{T}_{elim}$. Specifically, the structure of column $j$ of $L_{P}$ is obtained by combining the lower triangular sparsity structures of its children in $\mathcal{T}_{elim}$ with that of the corresponding column in $K_P^l(z_{\mu_ji}; p)$ \cite{scott2024cholesky}. During this process, indices are added to $\mathcal{S}_L$ in an order consistent with the dependencies encoded in $\mathcal{T}_{elim}$. Consequently, $\mathcal{S}_L$ forms an ordered set that dictates both the sparsity structure of $L_{P}$ and the column-processing order used during numerical factorization. The resulting sparsity structure of $L_P$ is illustrated in Figure \ref{fig:cholesky-exfactor}. Since symbolic factorization depends only on the sparsity pattern of $K_c^l(z_{\mu_ji};p)$, which remains fixed across interior-point and NMPC iterations, it is performed only once during the initial NMPC solve to compute $P$ and $\mathcal{S}_L$. Although symbolic factorization is a computational bottleneck in GPU-accelerated workflows \cite{pacaud2024gpu}, reusing these structures avoids repeating this cost in subsequent iterations.

Using the structural information obtained during symbolic factorization, the numerical factorization phase computes the numerical values of $L_P$. We denote this phase as:

\begin{equation} \label{eq:cholesky-numCholesky}
L_P = \text{NumCholesky}(K_c^l(z_{\mu_ji};p), P, \mathcal{S}_L)
\end{equation}

\noindent Here, the stored permutation $P$ is reused to access $K_P^l(z_{\mu_ji};p)$. Using the numerical values of $K_P^l(z_{\mu_ji};p)$, the entries $L_{P, ij}$ are computed for $(i, j) \in \mathcal{S}_L$ as follows \cite{scott2024cholesky}:

\begin{equation} \label{eq:cholesky-numeric}
L_{P, ij} =
\begin{cases}
\sqrt{K_{P,jj}^l - \displaystyle\sum_{\sigma=1}^{j-1} L_{P, j\sigma}^2}, & i=j \\[1.2em]
\dfrac{1}{L_{P, jj}}
\left(
K_{P,ij}^l -
\displaystyle\sum_{\sigma=1}^{j-1} L_{P, i\sigma}L_{P, j\sigma}
\right), & i>j
\end{cases}
\end{equation}

\noindent By restricting computations to the indices defined by $\mathcal{S}_L$ and following the column-processing order established during symbolic factorization, numerical factorization efficiently computes only the required non-zero entries of $L_P$. Unlike symbolic factorization, which is performed only once, numerical factorization is repeatedly performed to account for parametric changes in $K_c^l(z_{\mu_ji};p)$. As a result, reusing $P$ and $\mathcal{S}_L$ enables efficient repeated factorizations across interior-point (i.e., inner loop and outer loop) and NMPC iterations. For more details on sparse Cholesky algorithms, we refer the reader to \cite{davis2016sparse, scott2024cholesky}.

Since the factorization is performed on the reordered system $K_P^l(z_{\mu_{j}i}; p)$, we apply the same permutation to Equation \eqref{eq:liftedkkt-pd}. Defining the permuted primal search direction as $\Omega = P\delta_y^l$, the reordered linear system becomes:

\begin{equation} \label{eq:liftedkkt-permute}
L_PL_P^T\Omega = P(-r_1^{l} -G_i^{l}(\hat{y}_i; p)^T(Q_s^{l}r_4^{l} - r_2^{l}))
\end{equation}

\noindent Equation \eqref{eq:liftedkkt-permute} is solved via forward and backward substitution. First, the lower triangular system is solved for the intermediate vector $\upsilon$:

\begin{equation} \label{eq:cholesky-lower}
L_P\upsilon = P(-r^l_1 - G_i^l(\hat{y}_i; p)^T(Q^l_sr^l_4 - r^l_2))
\end{equation}

\noindent Next, we obtain $\Omega$ by solving the upper triangular system:

\begin{equation} \label{eq:cholesky-upper}
L_P^T \Omega = \upsilon
\end{equation}

\noindent Finally, we recover the original ordering of the search direction via:

\begin{equation} \label{eq:cholesky-delta}
\delta_y^l = P^T\Omega
\end{equation}

\noindent after which the other search directions can be solved for based on the equations given in \eqref{eq:infiniteOCP-condensed} and \eqref{eq:infiniteOCP-augmented}.

\subsubsection{Line Search and Loop Termination}
After computing the search directions $\delta_i$, we update $z_{\mu_j}$ using a line search with step size $\alpha$:

\begin{equation} \label{eq:infiniteOCP-lineSearch}
z_{\mu_ji+1} = z_{\mu_ji} + \alpha \delta_i
\end{equation}

\noindent We then proceed to the next inner loop iteration $i+1$, starting again from Eqn. \eqref{eq:infiniteOCP-augmented}. The inner loop terminates once the following condition is satisfied:

\begin{equation} \label{eq:infiniteOCP-innerTol}
E_{\mu_j}(z^*_{\mu_ji}) \leq \epsilon_{inner}
\end{equation}

\noindent where $E_{\mu_j}$ denotes the optimality error for the barrier problem for the fixed value $\mu_j$ and $\epsilon_{inner} > 0$ is the inner loop tolerance. Upon inner loop convergence, the barrier parameter is updated and the algorithm proceeds to the next outer loop iteration $j+1$ to solve a new barrier problem. Consequently, the algorithm repeatedly solves KKT systems with different numerical values but identical sparsity structure. The number of inner loop iterations required to compute $z^*_{\mu_{j+1}}$ can be reduced by initializing \eqref{eq:infiniteOCP-augmented} with the previous barrier solution  $z^*_{\mu_j}$, exploiting the fact that successive barrier solutions are expected to be close (i.e., $z^*_{\mu_{j+1}} \approx z^*_{\mu_j}$). The outer loop terminates once the following convergence criteria is satisfied:

\begin{equation} \label{eq:infiniteOCP-outerTol}
E_0(z^*_{\mu, k}) \leq \epsilon_{outer}
\end{equation}

\noindent where $z^*_{\mu, k}$ is the optimal solution to Problem \eqref{eq:infiniteOCP-general} at the control step $t_k$, $E_0$ is the optimality error evaluated at $\mu = 0$ and $\epsilon_{outer}$ is the outer loop tolerance. Similar to reducing inner loop iterations, the number of outer loop iterations within an NMPC solve can also be reduced via warmstarting, or initializing the solve for $z_{\mu, k+1}^*$ at the previous solution $z_{\mu, k}^*$. This improves the initial guess for interior-point, enabling the use of a smaller initial value for $\mu_j$ and accelerating algorithm convergence.

\subsubsection{Algorithm Summary}
Algorithm \ref{algo-nmpc} summarizes the proposed GPU-NMPC framework, illustrating the repeated cycle of solving $\mathcal{P}_k(p_k(t))$, applying control inputs to the system, and updating $p_k(t)$ using new system measurements at each $t_k$. Algorithm \ref{algo-interior} formalizes the proposed parametric interior-point method with a Lifted-KKT reformulation, denoted as LiftedIPM($\hat{\mathcal{P}}_k(\hat{p}_k), z_k, \mathcal{K}_{SIMD}, P, \mathcal{S}_L$). During the initial solve ($t_1 = 0$), SymCholesky($\cdot$) is performed once to obtain $P$ and $\mathcal{S}_L$. These structures are reused across interior-point iterations in Algorithm \ref{algo-interior} and NMPC iterations in Algorithm \ref{algo-nmpc}, amortizing the symbolic factorization cost across re-solves. In contrast, only numerical computations like NumCholesky($\cdot$) are recomputed to account for parametric changes in the KKT system. Together, these algorithms illustrate how the proposed framework exploits repeated problem structure to enable efficient GPU-accelerated NMPC re-solves.

\begin{algorithm}
\caption{GPU-accelerated NMPC workflow} \label{algo-nmpc}
\begin{algorithmic}[1]

\State \textbf{Given:} $\mathcal{T}$, $\mathcal{T}_p$, $\mathcal{T}_c$, initial guess $z_1$, $p_{1}(t)$
\State Define $\mathcal{P}_1(p_1(t))$
\State $\mathcal{K}_{SIMD}$, $\hat{\mathcal{P}}_1(\hat{p}_1) \leftarrow$ direct transcription using \eqref{eq:infinitesimd_form}
\State Initialize $P, \mathcal{S}_L$ = nothing

\For{$t_k$ in $\mathcal{T}$}
    \State $z^*_{\mu, k}, P, \mathcal{S}_L \leftarrow$ LiftedIPM($\hat{\mathcal{P}}_k(\hat{p}_k), z_k, \mathcal{K}_{SIMD}, P, \mathcal{S}_L$)
    \State Extract $y_c(\hat{t}_1)$ from $z^*_{\mu, k}$
    \State Apply $y_c(\hat{t}_1)$ to system actuator
    \State Obtain new state measurements $y_s(t_{k+1})$
    \State $z_{k+1} \leftarrow z_{\mu, k}^*$
    \State Update $\hat{p}_{k+1}$ from $p_k(t_{k+1})$ using $y_s(t_{k+1})$
\EndFor

\end{algorithmic}
\end{algorithm}

\begin{algorithm}
\caption{Interior-Point Method with Lifted-KKT (LiftedIPM)} \label{algo-interior}
\begin{algorithmic}[1]
\State \textbf{Given:} $\hat{\mathcal{P}}_k, z_k, \mathcal{K}_{SIMD}, P, \mathcal{S}_L$
\State $\hat{\mathcal{P}}^l_k$ $\leftarrow$ relax $\hat{h}(\cdot)$ with $\gamma$ using \eqref{eq:liftedkkt}
\State Initialize $z_{\mu} \leftarrow z_k$, $\mu \leftarrow \mu_0$
\While{\eqref{eq:infiniteOCP-outerTol} is not satisfied}
    \While{\eqref{eq:infiniteOCP-innerTol} is not satisfied}
        \State $K^l_{aug} \leftarrow$ initialize with $z_{\mu}, \mathcal{K}_{SIMD}$ using same procedure as \eqref{eq:infiniteOCP-augmented}
        \State $K^l_c \leftarrow$ condense $K_{aug}^l$ using \eqref{eq:liftedkkt-pd}
        \If{\eqref{eq:infiniteOCP-inertia} is not satisfied}
            \State Regularize with $\eta_y, \eta_c$ using same procedure as \eqref{eq:infiniteOCP-regularize} 
        \EndIf
        \If{$P, \mathcal{S}_L$ = nothing}
        \State $P, \mathcal{S}_L \leftarrow$ SymCholesky($K_c^l$)
        \EndIf
        \State $L_P\leftarrow$ NumCholesky($K_c^l, P, \mathcal{S}_L$)
        \State Evaluate $\delta_y^l$ by using $P, L_P$ to solve \eqref{eq:cholesky-lower}-\eqref{eq:cholesky-delta}
        \State Update $z_{\mu}$ with $\alpha, \delta_y^l$ using \eqref{eq:infiniteOCP-lineSearch}
    \EndWhile
    \State Update barrier parameter $\mu$
\EndWhile
    
\noindent \Return $z^*_{\mu}, P, \mathcal{S}_L$
\end{algorithmic}
\end{algorithm}

\subsection{Software Implementation} \label{sec:software}
The GPU-accelerated NMPC framework is implemented in the Julia packages \texttt{InfiniteOpt.jl} and \texttt{InfiniteExaModels.jl}. Initially, an \texttt{InfiniteModel} is constructed to define the initial OCP as defined in Problem \eqref{eq:infiniteOCP-continuous}. Problem parameters are specified via \texttt{@finite\_parameter} for finite values (e.g., initial conditions defined by instantaneous state measurements) and \texttt{@parameter\_function} for domain-dependent parameters (e.g., time-varying setpoints). The \texttt{InfiniteModel} is then transcribed using an \texttt{ExaTranscriptionBackend}, producing a finite, InfiniteSIMD-NLP \texttt{ExaModel} consistent with Problem \eqref{eq:infinitesimd_form}. While \texttt{ExaModel}s can be solved on CPU using JuliaSmoothOptimizers (JSO)-compliant solvers like \texttt{NLPModelsIpopt}, this work focuses on GPU execution using \texttt{MadNLP} with \texttt{cuDSS}. Specifically, \texttt{MadNLP} executes the overall interior-point algorithm, while \texttt{cuDSS} solves the linear system in Equation \eqref{eq:liftedkkt-pd} via sparse Cholesky factorization. During the initial factorization, symbolic factorization data such as $\mathcal{S}_L$ are constructed and cached within the solver instance for subsequent interior-point and NMPC iterations. Once the solution $z^*_{\mu}$ is computed, both the solution and solver instance are stored within the backend for reuse.

For subsequent re-solves, the framework supports in-place backend updates via two APIs. First, \texttt{warmstart\_backend\_start\_values} initializes the backend's primal and dual variables using the previous solution for warmstarted solves. Second, \texttt{set\_parameter\_value} updates the problem parameters without modifying the underlying problem structure. While either API may be used independently, only parameter updates preserve the existing backend and associated solver data. In contrast, warmstart-only workflows still require reconstruction of the solver backend at each iteration (requiring recomputation of the symbolic factorization) despite benefiting from improved initializations. These APIs are backend- and solver-agnostic, allowing the same update mechanism to be applied across different configurations, including CPU-based models via \texttt{JuMP.jl} or \texttt{ExaModels.jl} with \texttt{Ipopt} and GPU-based \texttt{ExaModel}s with \texttt{MadNLP}.

The user syntax for GPU-NMPC framework implementation is illustrated in Code Snippet \ref{code:exaGPU-resolves}. A more thorough discussion and tutorial on the implementation of the framework in \texttt{InfiniteOpt.jl} is available at:
\href{https://infiniteopt.github.io/InfiniteOpt.jl/dev/tutorials/resolves/}{https://infiniteopt.github.io/InfiniteOpt.jl/dev/tutorials/resolves/}.

\begin{figure}[!htp]
\begin{minipage}[t]{\linewidth}
\begin{scriptsize}
\lstset{language=Julia, breaklines = true}
\begin{lstlisting}[label = code:exaGPU-resolves, caption = Syntax for GPU-accelerated Re-solves in \texttt{InfiniteOpt.jl}.] 
using InfiniteOpt, InfiniteExaModels, MadNLPGPU, CUDA

# Define constants
dt = 1
y_init = [1.0]
tk = 0

# Simple function for simulating system state measurements
function simulate(y_state, u_opt, dt)
    dy = 4*sin(u_opt) - y_state^3
    return y_state + dt * dy
end

# Initialize the OCP to use InfiniteExaModels' GPU backend
model = InfiniteModel(ExaTranscriptionBackend(MadNLPSolver, backend = CUDABackend()))

# Define parameters & parameter functions
@infinite_parameter(model, t in [0, 10], num_supports = 15)
@finite_parameter(model, y0 == y_init[1])
@parameter_function(model, setpoint == t -> sin(t + tk))

# Define the problem constraints & objective as normal
@variable(model, 0 <= y, Infinite(t))
@variable(model, 1 <= u <= 3, Infinite(t))
@constraint(model, y(0) == y0)
@constraint(model, @deriv(y, t) == 4*sin(u) - y^3)
@objective(model, Min, @integral((y - setpoint)^2, t))
    
# Define the NMPC loop
for tk in 1:dt:2
    # Solve the model
    optimize!(model)

    # Apply optimized control input to get new state measurements
    u_opt = @CUDA.allowscalar(value(u)[2])
    y_init[1] = simulate(y_init[1], u_opt, dt)
    
    # Warmstarting
    warmstart_backend_start_values(model)
    
    # Updating parameter values
    set_parameter_value(y0, y_init[1])
    setpoint_new = (t) -> sin(t + tk)
    set_parameter_value(setpoint, setpoint_new)
end

\end{lstlisting}
\end{scriptsize}
\end{minipage}
\end{figure}

\section{Case Studies} \label{sec:caseStudies}
In the following section, we demonstrate the capabilities of the InfiniteOpt GPU-NMPC framework with representative NMPC case studies in dynamic and PDE-constrained optimization. CPU-based configurations consisting of \texttt{JuMP.jl} and \texttt{InfiniteExaModels.jl} are solved with \texttt{Ipopt} v3.14.19 and the HSL linear solver \texttt{MA97} v2.8.1. For \texttt{JuMP.jl}, we consider both the default reverse-mode AD and symbolic AD in the \texttt{Nonlinear.SymbolicAD} submodule of \texttt{MathOptInterface.jl}. GPU configurations with \texttt{InfiniteExaModels.jl} and \texttt{OptimalControl.jl} are solved using \texttt{MadNLP.jl} v0.8.12 and \texttt{cuDSS} v0.7.1. For solving with the CUDA C/C++ based solver \texttt{MPCGPU}, we use \texttt{CUDA} v13.0 and compile with \texttt{g++} v13.3. Other package versions include \texttt{InfiniteOpt.jl} v0.6.3, \texttt{JuMP.jl} v1.30.0, \texttt{MathOptInterface.jl} v1.49.0, \texttt{InfiniteExaModels.jl} v0.1.2 and \texttt{OptimalControl.jl} v1.1.6. The benchmark results are collected using a Linux machine running Ubuntu with an Intel(R) Xeon(R) w5-3435X CPU @ 3.10GHz, 512GB of memory, and a RTX 6000 Ada Generation GPU with 48GB of VRAM. The results have been averaged over three independent runs. The source code for the case studies is available at \url{https://github.com/infiniteopt/GPU-NMPC-paper}.

\subsection{Distillation Column}
Taking inspiration from \cite{pacaud2024gpu}, we aim to control a distillation column with $N$ trays. The feed is sent to tray $N_f$ with flow rate $F$ and concentration $x_f$, We also fix the distillate flow rate $D$. The objective is to achieve the steady-state values $\bar{x}$ and $\bar{u}$, which are for the tray 1 liquid mole fraction and reflux ratio, respectively. The MPC simulation is set from $t_0$ = 0 s to $t_f$ = 300 s, with prediction and control horizons $\mathcal{T}_p$ = $\mathcal{T}_c$ = 180 s and $\Delta t$ = 2 s. Backward finite difference was chosen for the derivative approximation method. The problem parameters and OCP formulation are given in Table \ref{distill-params} and Problem \eqref{eq:distill-ocp}, respectively:

\begin{table}[!htb]
\caption{Distillation Benchmark Parameters} \label{distill-params}
\begin{adjustbox}{center=\textwidth}
\begin{tabular}{|c|c|c|c|c|c|c|c|c|c|c|}
    \hline
    $N$ & $N_f$ & $D$ & $F$ & $\alpha$ & $\bar{x}$ & $\bar{u}$ & $x_f$ & $A_c$ & $A_t$ & $A_r$ \\
    \hline
    32 & 17 & 0.2 & 0.4 & 1.6 & 0.8958 & 2.51459 & 0.5 & 0.5 & 0.25 & 1.0 \\
    \hline
\end{tabular}
\end{adjustbox}
\end{table}

\begin{figure}[!htb]
     \centering
     \begin{subfigure}[b]{0.49\textwidth}
         \centering
         \includegraphics[width=\textwidth]{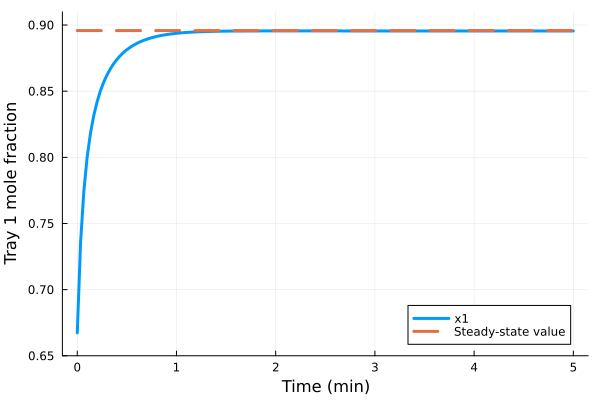}
         \caption{Tray 1 Liquid Mole Fraction over Time}
         \label{fig:distill-x1}
     \end{subfigure}
     \hfill
     \begin{subfigure}[b]{0.49\textwidth}
         \centering
         \includegraphics[width=\textwidth]{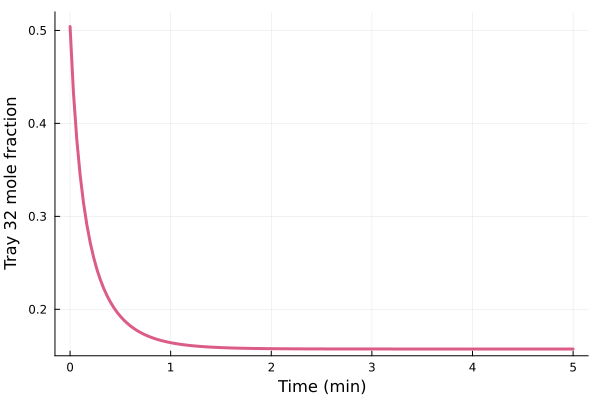}
         \caption{Tray 32 Vapor Mole Fraction over Time}
         \label{fig:distill-trayfrac}
     \end{subfigure}
    \caption{Tray 1 Liquid and Tray 32 Vapor Mole Fractions over Time For Distillation Benchmark}
    \label{fig:distill-x1-y32}
\end{figure}

\begin{figure}[!htb]
     \centering
     \begin{subfigure}[b]{0.49\textwidth}
         \centering
         \includegraphics[width=\textwidth]{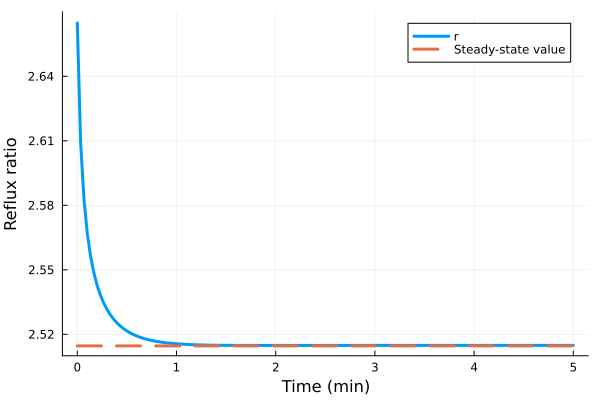}
         \caption{Optimal reflux ratio input over time}
         \label{fig:distill-reflux}
     \end{subfigure}
     \hfill
     \begin{subfigure}[b]{0.49\textwidth}
         \centering
         \includegraphics[width=\textwidth]{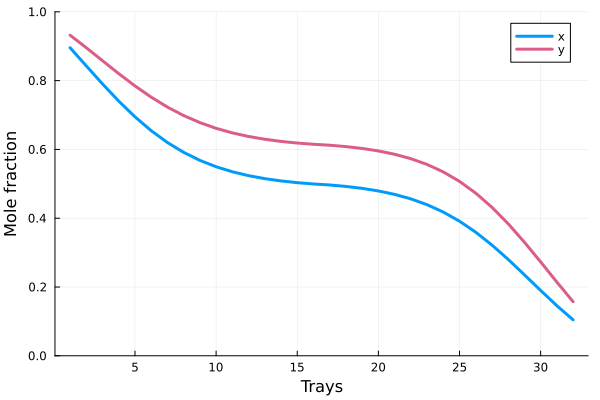}
         \caption{Liquid (x) and vapor (y) tray mole fractions}
         \label{fig:distill-trayfrac}
     \end{subfigure}
    \caption{Reflux ratio over time and mole fractions across all trays in the distillation benchmark}
    \label{fig:distill-reflux-trayfrac}
\end{figure}

The evolution of the first tray liquid mole fraction and final tray vapor mole fractions over time are shown in Figure \ref{fig:distill-x1-y32}, while the reflux ratio and the liquid-vapor mole fractions across all trays are shown in Figure \ref{fig:distill-reflux-trayfrac}. Figures \ref{fig:distill-x1} and \ref{fig:distill-reflux} show that the trajectories reach $\bar{x}$ and $\bar{u}$ after approximately 1 minute, after which the system exhibits stable closed-loop behavior over the remainder of the NMPC horizon.

\begin{equation} \label{eq:distill-ocp}
\begin{aligned}
& \min_{x(t), u(t)} && \int_{0}^{T_p} (x_1(t)-\bar{x})^2+0.1(u(t)-\bar{u})^2dt\\\
& \textrm{s.t.} & \\
&&& x_{i}(0)=x_{0} & i \in \{1, 2,..., N\}\\
&&& y_{i}(0)=y_0 & i \in \{1, 2,..., N\}\\
&&&y_{i}(t) = \frac{\alpha x_i(t)}{1+(\alpha-1)x_i(t)} & i \in \{1, 2,...,N\}\\
&&& L(t) = u(t)D \\
&&& V(t) = L(t)+D \\
&&& S(t) = L(t)+F \\
&&&\frac{dx_1(t)}{dt}=\frac{V(t)(y_2(t) - x_1(t))}{A_c}\\
&&&\frac{dx_j(t)}{dt}=\frac{L(t)(x_{j-1}(t) -  x_{j}(t)) - V(t)(y_{j}(t) - y_{j+1}(t))}{At} & j \in \{2, 3,..., N_{f}-1\} \\
&&&\frac{dx_{N_{f}}(t)}{dt}=\frac{Fx_f + L(t)x_{N_{f}-1}(t) - S(t)x_{N_f}(t) - V(t)(y_{N_f}(t) - y_{N_f+1}(t))}{At}\\
&&&\frac{dx_k(t)}{dt}=\frac{S(t)(x_{k-1}(t) - x_{k}(t)) - V(t)(y_{k}(t) - y_{k+1}(t))}{At} & k \in \{N_f+1,..., N\}\\
&&&\frac{dx_N(t)}{dt}=\frac{S(t)x_{N-1}(t) - (F - D)x_N(t) - V(t)y_N(t))}{Ar}\\
&&& 0\leq x_i(t) \leq 1 & i \in \{1, 2, ..., N\} \\
&&& 0\leq y_i(t) \leq 1 & i \in \{1, 2, ..., N\} \\
&&& 1\leq u(t) \leq 5\\
&&& 0\leq L(t) \leq 10\\
&&& 0\leq V(t) \leq 10\\
&&& 0\leq S(t) \leq 10\\
\end{aligned}
\end{equation}

\subsection{PDE Heated Plate}
An aluminum plate is defined over the domain \(x \in [-1,1] \times [-1,1]\) with thermal conductivity $\alpha$, length $L$, fixed boundary temperature of $T_b$ and heat loss $q_{loss}$. Each spatial dimension is discretized with $n$ points, creating a grid of $n \times n$ points. The objective is to control the average temperature within a specified region by manipulating the input $Q(t)$, which heats the left and bottom sides of the plate. Spatial derivatives are transcribed using second-order central finite difference and the time derivatives are approximated using backward finite difference. The MPC simulation is set with $t_f$ = 3.0 s, $\mathcal{T}_p$ = $\mathcal{T}_c$ = 1.0 s and $\Delta t$ = 0.08 s. The problem parameters and OCP formulation are given in Table \ref{pde-params} and Problem \eqref{eq:pde-ocp}, respectively:

\begin{table}[H]
\caption{PDE Heated Plate Benchmark Parameters} \label{pde-params}
\begin{adjustbox}{center=\textwidth}
\begin{tabular}{|c|c|c|c|c|c|c|c|c|c|c|}
    \hline
    $\alpha$ & $L$ & $T_{b}$ & n & $\rho$ & $q_{loss}$ \\
    \hline
    230 & 0.1 & 50.0 & 21 & $1 \times 10^{-8}$ & 10 \\
    \hline
\end{tabular}
\end{adjustbox}
\end{table}

\begin{equation} \label{eq:pde-ocp}
\begin{aligned}
& \min_{T(t, x_2, x_2), Q(t)} && \int_{0}^{T_p} \left( \iint_{-0.5}^{0.5} (T(t, x_1, x_2) - T_{\text{sp}}(t))^2 \, dx_1\, dx_2  \right) + \rho\, {Q(t)}^2 dt \\
& \textrm{s.t.} && T_{sp}(t) = 
\left\{
\begin{aligned}
&& 58, & \quad 0 \leq t < 0.5 \\
&& 55, & \quad 0.5 \leq t < 1.5 \\
&& 63, & \quad 1.5 \leq t \leq t_f \\
\end{aligned}
\right. \\
&&& T(0, x_1, x_2) = T_0 \\
&&&\frac{dT(t, x_1, x_2)}{dt}=\alpha\left(\frac{d^2T(t, x_1, x_2)}{dx_1^2}+\frac{d^2T(t, x_1, x_2)}{dx_2^2} - 5q_{loss}^2\right )\\
&&&T(t, -1, x_2) = 4.1\sqrt[3]{Q(t)} \\
&&&T(t, 1, x_2) = T_b \\
&&&T(t, x_1, -1) = \sqrt{Q(t)} \\
&&&T(t, x_1, 1) = T_b \\
&&& 50 \leq Q(t) \\
&&& x_1 \in [-1, 1] \\
&&& x_2 \in [-1, 1] \\
&&& t \in [0, T_p]
\end{aligned}
\end{equation}

\begin{figure}[!htb]
     \centering
     \begin{subfigure}[b]{0.49\textwidth}
         \centering
         \includegraphics[width=\textwidth]{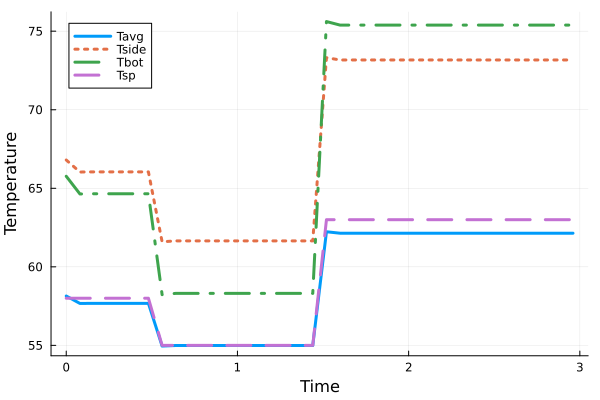}
         \caption{The Average Region (Tavg), Side (Tside), Bottom (Tbot) and Setpoint (Tsp) Temperatures over Time}
         \label{fig:pde-temp}
     \end{subfigure}
     \hfill
     \begin{subfigure}[b]{0.49\textwidth}
         \centering
         \includegraphics[width=\textwidth]{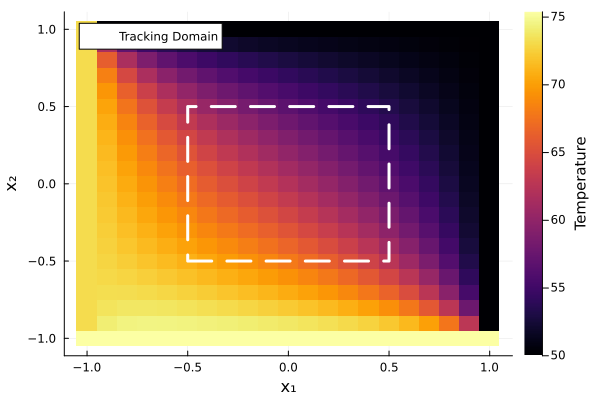}
         \caption{Final Temperature Distribution Across Heated Plate at Time $t_f$}
         \label{fig:pde-distrib}
     \end{subfigure}
    \caption{Temperature Tracking and Final Temperature Distribution for NMPC Control of a 2D Heated Plate Region\protect}
    \label{fig:pde-results}
\end{figure}

The NMPC results are shown in Figure \ref{fig:pde-results}. Figure \ref{fig:pde-distrib} illustrates the final temperature distribution of the heated plate together with the tracking region used to compute the average temperature for setpoint tracking. Figure \ref{fig:pde-results} also shows that the controller successfully reaches the setpoint temperature of 55, while only closely approaching the higher setpoint temperatures.

\subsection{Benchmarking Results and Discussion}
\begin{table}[!htb]
\caption{Average Total NMPC Simulation Times (s) over CPU and GPU Configurations Under Different Warmstart (WS) and Parameter Update (PU) Settings for the Distillation and PDE Plate Benchmarks} \label{nmpc-total-times}
\begin{adjustbox}{center=\textwidth}
\begin{tabular}{|c|c|c|c|}
    \hline
    \multicolumn{2}{|c|}{\textbf{Configuration}} & \textbf{Distillation} & \textbf{PDE Plate}\\
    \hline
    \multirow{4}{*}{\textbf{{\thead{JuMP +\\ Reverse-mode AD +\\ Ipopt}}}} & \textbf{Base} & 58.72 & 125.41\\
    & \textbf{WS} & 51.62 & 88.20 \\
    & \textbf{PU} & 24.52 & 119.11 \\
    & \textbf{WS + PU} & 19.06 & 77.33 \\
    \hline
    \multirow{4}{*}{\textbf{{\thead{JuMP +\\ Symbolic AD +\\ Ipopt}}}} & \textbf{Base} & 47.78 & 108.34 \\
    & \textbf{WS} & 44.82 & 85.41 \\
    & \textbf{PU} & 30.75 & 96.02 \\
    & \textbf{WS + PU} & 17.96 & 74.51 \\
    \hline
    \multirow{4}{*}{\textbf{\thead{InfiniteExaModels +\\ NLPModelsIpopt}}} & \textbf{Base} & 22.81 & 86.25 \\
    & \textbf{WS} & 22.92 & 67.57 \\
    & \textbf{PU} & 21.34 & 84.42 \\
    & \textbf{WS + PU} & 8.80 & 36.50 \\
    \hline
    \multirow{4}{*}{\textbf{\thead{InfiniteExaModels +\\ MadNLP +\\ cuDSS}}} & \textbf{Base} & 21.91 & 37.30 \\
    & \textbf{WS} & 37.76 & 54.75 \\
    & \textbf{PU} & 11.77 & 31.63 \\
    & \textbf{WS + PU} & 3.00 & 3.28 \\
    \hline
    \multirow{2}{*}{ \textbf{\thead{OptimalControl +\\ ExaModels +\\ MadNLP + cuDSS}}} & 
    \rule[-2.5ex]{0pt}{5.0ex} \textbf{Base} & 20.14 & N/A \\
    & \rule[-2.0ex]{0pt}{3.0ex} \textbf{WS} & 15.13 & N/A \\
    \hline
    \textbf{\thead{MPCGPU}} & \textbf{WS} & 2.44 & N/A \\
    \hline
\end{tabular}
\end{adjustbox}
\end{table}

Table \ref{nmpc-total-times} reports the total NMPC simulation times for both case studies, comparing CPU and GPU configurations under different warmstarting (WS) and parameter update (PU) settings. Note that the \texttt{OptimalControl.jl} and \texttt{MPCGPU} configurations were only evaluated on the distillation benchmark, as neither package provides native support for PDEs. Furthermore, \texttt{OptimalControl.jl} does not support parameter updates to the best of our knowledge. We also note that \texttt{MPCGPU} does not provide a high-level modeling API, but rather requires the user to provide the necessary function and gradient evaluations based on manual transcription in CUDA C/C++. Thus, there is no notion of model creation or AD, so the reported times are only based on solver times. Since the baseline and warmstarted \texttt{MPCGPU} configurations exhibited nearly identical timings, only the warmstarted results are reported.

Across the CPU-based workflows, the WS + PU configuration reduces NMPC times by up to 67\% and 58\% compared to the base case for the distillation and PDE benchmarks, respectively. These improvements largely stem from avoiding repeated model construction and faster solver convergence through warmstarting, highlighting the benefits of backend reuse on CPU.

For the GPU-based workflows, there are significant improvements with \texttt{InfiniteExaModels.jl} where the WS + PU configuration reduces NMPC times by over an order-of-magnitude relative to the base case. In contrast, the WS-only configuration yields the worst overall NMPC times among the four configurations. Based on Tables \ref{distill-bd} and \ref{pde-bd} in the Appendix, this primarily stems from substantially higher setup costs of warmstart initialization, even though the solve times are comparable to or faster than those of the base case. This reflects an implementation tradeoff, as the framework is optimized for the combined WS + PU workflow. The \texttt{OptimalControl.jl} GPU configuration times are comparable to the WS + PU \texttt{JuMP} workflows, indicating that even with GPU acceleration, there is substantial overhead from reconstructing the backend and the symbolic factorization at each NMPC iteration. This is especially evident for the PDE benchmark in Table \ref{pde-bd} from the Appendix, where up to 96\% of the per-iteration solve times is attributed to the linear solver, which includes the factorization routine.

In comparison, the \texttt{MPCGPU} workflow performs slightly faster than \texttt{InfiniteExaModels.jl}. Since the distillation benchmark is manually hard-coded within the solver, there is no model construction overhead nor AD. On the other hand, \texttt{InfiniteExaModels.jl} automatically performs transcription and differentiation from a high-level problem formulation, trading some computational efficiency for a more general and user-friendly modeling interface. We also note that while the \texttt{MPCGPU} and Julia-based configurations produce nearly identical state trajectories, the \texttt{MPCGPU} control input trajectory remains at the steady-state value $\bar{u}$ throughout the NMPC simulation, whereas the Julia-based trajectories match that of Figure \ref{fig:distill-reflux}. This discrepency is likely due to differences in the underlying numerical formulations and solver algorithms implemented in \texttt{MPCGPU} and the Julia-based workflows.

\begin{figure}[!htb]
     \centering
     \begin{subfigure}[b]{0.52\textwidth}
         \centering
         \includegraphics[width=\textwidth]{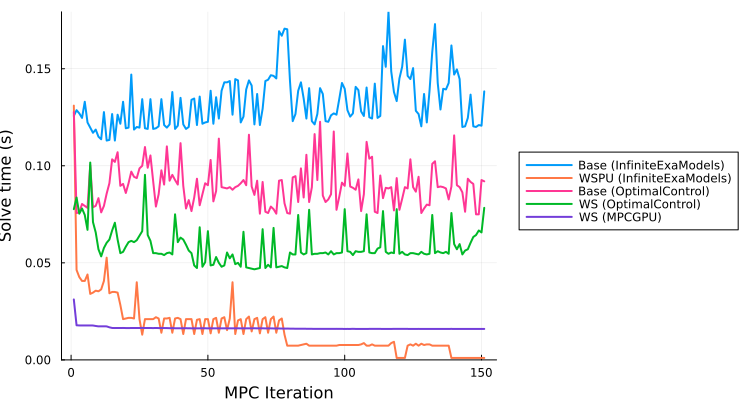}
         \caption{Distillation benchmark}
         \label{fig:distill-gpusolveits-combined}
     \end{subfigure}
     \hfill
     \begin{subfigure}[b]{0.41\textwidth}
         \centering
         \includegraphics[width=\textwidth]{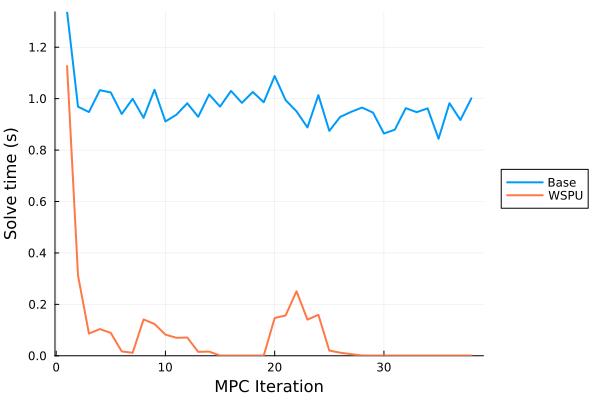}
         \caption{PDE plate benchmark}
         \label{fig:pde-gpusolveits-combined}
     \end{subfigure}
    \caption{Average GPU Solve times per NMPC Iteration for the Distillation and PDE Plate Benchmarks with the InfiniteExaModels, OptimalControl and MPCGPU Configurations}
    \label{fig:benchmark-gpusolveits-combined}
\end{figure}

Figure \ref{fig:benchmark-gpusolveits-combined} illustrates average per-iteration GPU solve times across all NMPC iterations for both benchmarks. \texttt{InfiniteExaModels} achieves reductions of 85\% and 94\% compared to the base case for the distillation and PDE benchmarks, respectively. The initial spike reflects the high initialization cost of the first solve, showcasing the computational overhead associated with the initial factorization and lack of a warmstart. Subsequent iterations exhibit significantly lower solve times. For the PDE benchmark, the solve time spikes around iterations 8 and 20 in Figure \ref{fig:pde-gpusolveits-combined} coincide with changes in the setpoint temperature. Similar trends are observed in the corresponding CPU benchmark figures in the Appendix (\ref{fig:pde-ipoptsolveits-combined}, \ref{fig:pde-moisolveits-combined}, and \ref{fig:pde-cpusolveits-combined}), reflecting the reduced effectiveness of WS under substantial setpoint changes. While the \texttt{OptimalControl} configuration indicates that WS alone reduces solve times considerably, combining it with factorization reuse leads to additional speedups. This is evident with the \texttt{InfiniteExaModels} WS + PU configuration, which combines both effects and performs similarily to \texttt{MPCGPU}. Similar solve time trends are observed for the CPU-based \texttt{JuMP} and \texttt{InfiniteExaModels} configurations in Figures \ref{fig:benchmark-ipoptsolveits-combined}, \ref{fig:benchmark-moisolveits-combined} and \ref{fig:benchmark-cpusolveits-combined} from the Appendix, showcasing that performance improvements are not limited to GPU workflows.

\section{Conclusions and Future Work} 
\label{sec:conclusion}
We have introduced a second-order, gradient-based direct GPU-NMPC framework implemented in \texttt{InfiniteOpt.jl} and \texttt{InfiniteExaModels.jl}. Using a parametric interior-point formulation, the framework exploits the fixed structure of OCPs to reuse structure-dependent computations such as symbolic Cholesky factorization to amortize initialization overhead in NMPC workflows. Moreover, the recurrent algebraic structure of transcribed OCPs is leveraged through the InfiniteSIMD-NLP abstraction, enabling efficient parallelized modeling and solution of NMPC problems on GPUs. Results on distillation and PDE heated plate benchmarks demonstrate up to 94\% reductions in per-NMPC iteration GPU solve times and over an order-of-magnitude speedup in total NMPC times. These results highlight the effectiveness of reusing problem structure and previous solution information, demonstrating the potential of the proposed approach to improve the feasiblity of real-time NMPC applications.

As a future direction, linear solver performance may be further improved through tailored constraint reordering strategies to produce KKT structures that are more amenable to efficient sparse factorization. Moreover, as the methods behind the proposed framework are general, it may also be extended beyond NMPC to other re-solve applications, such as multi-objective optimization.

\section*{Acknowledgements}
We acknowledge financial support from the Natural Sciences and Engineering Research Council of Canada under RGPIN-2024-03997 and the Faculty of Engineering at the University of Waterloo.

\bibliography{refs}

@article{pulsipher2022unifying,
    title={{A unifying modeling abstraction for infinite-dimensional optimization}},
    author={Pulsipher, Joshua L. and Zhang, Weiqi and Hongisto, Tyler J and Zavala, Victor M},
    journal={Computers \& Chemical Engineering},
    volume={156},
    pages={107567},
    year={2022},
    publisher={Elsevier},
}

@incollection{pulsipher2024scalable,
    title={{Scalable Modeling of Infinite-Dimensional Nonlinear Programs with InfiniteExaModels. jl}},
    author={Pulsipher, Joshua L. and Shin, Sungho},
    booktitle={Computer Aided Chemical Engineering},
    volume={53},
    pages={3373--3378},
    year={2024},
    publisher={Elsevier},
}

@misc{shin2024opf,
      title={{Accelerating Optimal Power Flow with GPUs: SIMD Abstraction of Nonlinear Programs and Condensed-Space Interior-Point Methods}}, 
      author={Sungho Shin and François Pacaud and Mihai Anitescu},
      year={2024},
      eprint={2307.16830},
      archivePrefix={arXiv},
      primaryClass={math.OC},
      url={https://arxiv.org/abs/2307.16830}, 
}

@misc{pacaud2024gpu,
      title={{GPU-accelerated dynamic nonlinear optimization with ExaModels and MadNLP}}, 
      author={François Pacaud and Sungho Shin},
      year={2024},
      eprint={2403.15913},
      archivePrefix={arXiv},
      primaryClass={math.OC},
      url={https://arxiv.org/abs/2403.15913}, 
}

@article{gondosiswanto2025advances,
    title = {{Advances to modeling and solving infinite-dimensional optimization problems in InfiniteOpt.jl}},
    journal = {Digital Chemical Engineering},
    pages = {100236},
    year = {2025},
    issn = {2772-5081},
    doi = {https://doi.org/10.1016/j.dche.2025.100236},
    url = {https://www.sciencedirect.com/science/article/pii/S2772508125000201},
    author = {Evelyn Gondosiswanto and Joshua L. Pulsipher},
}

@inproceedings{adabag2024gpu,
    title={{MPCGPU: Real-Time Nonlinear Model Predictive Control through Preconditioned Conjugate Gradient on the GPU}},
    url={http://dx.doi.org/10.1109/ICRA57147.2024.10611212},
    DOI={10.1109/icra57147.2024.10611212},
    booktitle={2024 IEEE International Conference on Robotics and Automation (ICRA)},
    publisher={IEEE},
    author={Adabag, Emre and Atal, Miloni and Gerard, William and Plancher, Brian},
    year={2024},
    month=may,
    pages={9787–9794}
}

@ARTICLE{yurtsever2020auto,
  author={Yurtsever, Ekim and Lambert, Jacob and Carballo, Alexander and Takeda, Kazuya},
  journal={IEEE Access}, 
  title={{A Survey of Autonomous Driving: Common Practices and Emerging Technologies}}, 
  year={2020},
  volume={8},
  number={},
  pages={58443-58469},
  doi={10.1109/ACCESS.2020.2983149},
}

@article{eren2017aerospace,
    title = {{"Model Predictive Control in Aerospace Systems: Current State and Opportunities"}},
    author = "Utku Eren and Anna Prach and Ko{\c c}er, \{Basaran Bahadir\} and Rakovic, \{Sa{\v s}a V.\} and Erdal Kayacan and Beh{\c c}et A{\c c}ikmese",
    year = "2017",
    doi = "10.2514/1.G002507",
    language = "English",
    volume = "40",
    pages = "1541--1566",
    journal = "Journal of Guidance, Control, and Dynamics",
    issn = "0731-5090",
    publisher = "American Institute of Aeronautics and Astronautics Inc. (AIAA)",
    number = "7",
}

@InProceedings{qin2000nmpc,
    author={Qin, S. Joe and Badgwell, Thomas A.},
    title={{An Overview of Nonlinear Model Predictive Control Applications}},
    booktitle={Nonlinear Model Predictive Control},
    bookauthor={Allg{\"o}wer, Frank and Zheng, Alex},
    journal={Progress in Systems and Control Theory},
    year={2000},
    publisher={Birkh{\"a}user Basel},
    pages={369-392},
    isbn={978-3-0348-8407-5},
}

@article{taheri2022hvac,
    title={{Model predictive control of heating, ventilation, and air conditioning (HVAC) systems: A state-of-the-art review.}},
    author={Taheri, Saman and Hosseini, Paniz and Razban, Ali},
    journal={Journal of Building Engineering},
    volume={60},
    year={2022},
}

@article{allgower2004nmpc,
    title={{Nonlinear model predictive control: From theory to application}},
    author={Allg{\"{o}}wer, Frank and Findeisen, Rolf and Nagy, Zoltan K.},
    journal={Journal of the Chinese Institute of Chemical Engineers},
    volume={35},
    issue={3},
    pages={299-315},
    year={2004}
}

@article{marti2013distribute,
    title = {{A method to coordinate decentralized NMPC controllers in oxygen distribution networks}},
    journal = {Computers \& Chemical Engineering},
    volume = {59},
    pages = {122-137},
    year = {2013},
    note = {Selected papers from ESCAPE-22 (European Symposium on Computer Aided Process Engineering - 22), 17-20 June 2012, London, UK},
    issn = {0098-1354},
    doi = {https://doi.org/10.1016/j.compchemeng.2013.05.023},
    url = {https://www.sciencedirect.com/science/article/pii/S0098135413001877},
    author = {Rubén Martí and Daniel Sarabia and Daniel Navia and César {de Prada}},
}

@article{liu2009distribute,
    title = {{Distributed Model Predictive Control of Nonlinear Process Systems Subject to Asynchronous Measurements}},
    journal = {IFAC Proceedings Volumes},
    volume = {42},
    number = {11},
    pages = {147-152},
    year = {2009},
    note = {7th IFAC Symposium on Advanced Control of Chemical Processes},
    issn = {1474-6670},
    doi = {https://doi.org/10.3182/20090712-4-TR-2008.00021},
    url = {https://www.sciencedirect.com/science/article/pii/S1474667015302640},
    author = {Jinfeng Liu and David Muñoz {de la Peña} and Panagiotis D. Christofides},
}

@article{zhu2000decomposition,
    title = {{A hybrid model predictive control strategy for nonlinear plant-wide control}},
    journal = {Journal of Process Control},
    volume = {10},
    number = {5},
    pages = {449-458},
    year = {2000},
    issn = {0959-1524},
    doi = {https://doi.org/10.1016/S0959-1524(00)00020-2},
    url = {https://www.sciencedirect.com/science/article/pii/S0959152400000202},
    author = {Guang-Yan Zhu and Michael A. Henson and Babatunde A. Ogunnaike},
}

@article{bradley2022perspectives,
    title = {{Perspectives on the integration between first-principles and data-driven modeling}},
    journal = {Computers \& Chemical Engineering},
    volume = {166},
    pages = {107898},
    year = {2022},
    issn = {0098-1354},
    doi = {https://doi.org/10.1016/j.compchemeng.2022.107898},
    url = {https://www.sciencedirect.com/science/article/pii/S0098135422002368},
    author = {William Bradley and Jinhyeun Kim and Zachary Kilwein and Logan Blakely and Michael Eydenberg and Jordan Jalvin and Carl Laird and Fani Boukouvala},
}

@article{shah2025hybrid,
    title = {{Hybrid modeling of first-principles and machine learning: A step-by-step tutorial review for practical implementation}},
    journal = {Computers \& Chemical Engineering},
    volume = {194},
    pages = {108926},
    year = {2025},
    issn = {0098-1354},
    doi = {https://doi.org/10.1016/j.compchemeng.2024.108926},
    url = {https://www.sciencedirect.com/science/article/pii/S0098135424003442},
    author = {Parth Shah and Silabrata Pahari and Raj Bhavsar and Joseph Sang-Il Kwon},
}

@ARTICLE{piche2000neural,
  author={Piche, S. and Sayyar-Rodsari, B. and Johnson, D. and Gerules, M.},
  journal={IEEE Control Systems Magazine}, 
  title={{Nonlinear model predictive control using neural networks}}, 
  year={2000},
  volume={20},
  number={3},
  pages={53-62},
  doi={10.1109/37.845038},
}

@article{luo2023node,
    title = {{Model predictive control of nonlinear processes using neural ordinary differential equation models}},
    journal = {Computers \& Chemical Engineering},
    volume = {178},
    pages = {108367},
    year = {2023},
    issn = {0098-1354},
    doi = {https://doi.org/10.1016/j.compchemeng.2023.108367},
    url = {https://www.sciencedirect.com/science/article/pii/S0098135423002375},
    author = {Junwei Luo and Fahim Abdullah and Panagiotis D. Christofides},
}

@article{casas2025pinn,
    title = {{A comparison of strategies to embed physics-informed neural networks in nonlinear model predictive control formulations solved via direct transcription}},
    journal = {Computers \& Chemical Engineering},
    volume = {198},
    pages = {109105},
    year = {2025},
    issn = {0098-1354},
    doi = {https://doi.org/10.1016/j.compchemeng.2025.109105},
    url = {https://www.sciencedirect.com/science/article/pii/S0098135425001097},
    author = {Carlos Andrés {Elorza Casas} and Luis A. Ricardez-Sandoval and Joshua L. Pulsipher},
}

@article{antonelo2024pinn,
   title={{Physics-informed neural nets for control of dynamical systems}},
   volume={579},
   ISSN={0925-2312},
   url={http://dx.doi.org/10.1016/j.neucom.2024.127419},
   DOI={10.1016/j.neucom.2024.127419},
   journal={Neurocomputing},
   publisher={Elsevier BV},
   author={Antonelo, Eric Aislan and Camponogara, Eduardo and Seman, Laio Oriel and Jordanou, Jean Panaioti and de Souza, Eduardo Rehbein and Hübner, Jomi Fred},
   year={2024},
   month=apr,
   pages={127419},
}

@misc{montoison2025ocp,
  title={{Modeling and Optimization of Control Problems on GPUs}}, 
  author={Alexis Montoison and Jean-Baptiste Caillau},
  year={2025},
  eprint={2510.03932},
  archivePrefix={arXiv},
  primaryClass={math.OC},
  url={https://arxiv.org/abs/2510.03932}, 
}

@article{nicholson2018pyomo,
  author    = {Bethany Nicholson and John D. Siirola and Jean{-}Paul Watson and Victor M. Zavala and Lorenz T. Biegler},
  title     = {{pyomo.dae: A Modeling and Automatic Discretization Framework for Optimization with Differential and Algebraic Equations}},
  journal   = {Mathematical Programming Computation},
  year      = {2018},
  volume    = {10},
  number    = {2},
  pages     = {187--223},
  doi       = {10.1007/s12532-017-0127-0},
  url       = {https://doi.org/10.1007/s12532-017-0127-0},
  issn      = {1867-2957}
}

@article{wachter2006ipopt,
  author  = {W{\"a}chter, Andreas and Biegler, Lorenz T.},
  title   = {{On the implementation of an interior-point filter line-search algorithm for large-scale nonlinear programming}},
  journal = {Mathematical Programming},
  volume  = {106},
  number  = {1},
  pages   = {25--57},
  year    = {2006},
  doi     = {10.1007/s10107-004-0559-y},
  url     = {https://doi.org/10.1007/s10107-004-0559-y},
  issn    = {1436-4646},
}

@article{davis2016sparse,
title={A survey of direct methods for sparse linear systems}, volume={25},
DOI={10.1017/S0962492916000076},
journal={Acta Numerica},
author={Davis, Timothy A. and Rajamanickam, Sivasankaran and Sid-Lakhdar, Wissam M.},
year={2016},
pages={383–566},
}

@article{margossian2019ad,
   title={{A review of automatic differentiation and its efficient implementation}},
   volume={9},
   ISSN={1942-4795},
   url={http://dx.doi.org/10.1002/WIDM.1305},
   DOI={10.1002/widm.1305},
   number={4},
   journal={WIREs Data Mining and Knowledge Discovery},
   publisher={Wiley},
   author={Margossian, Charles C.},
   year={2019},
   month={03},
}

@article{dunning2017jump,
   title={{JuMP: A Modeling Language for Mathematical Optimization}},
   volume={59},
   ISSN={1095-7200},
   url={http://dx.doi.org/10.1137/15M1020575},
   DOI={10.1137/15m1020575},
   number={2},
   journal={SIAM Review},
   publisher={Society for Industrial & Applied Mathematics (SIAM)},
   author={Dunning, Iain and Huchette, Joey and Lubin, Miles},
   year={2017},
   month=jan, pages={295–320}
   }

@article{gay2015ampl,
    author = {Gay, David M.},
    year = {2015},
    month = {01},
    pages = {95-116},
    journal = {Numerical Analysis and Optimization},
    publisher={Springer International Publishing},
    title = {{The AMPL Modeling Language: An Aid to Formulating and Solving Optimization Problems}},
    isbn = {978-3-319-17688-8},
    doi = {10.1007/978-3-319-17689-5_5}
}

@misc{falanga2018mpc,
      title={{PAMPC: Perception-Aware Model Predictive Control for Quadrotors}}, 
      author={Davide Falanga and Philipp Foehn and Peng Lu and Davide Scaramuzza},
      year={2018},
      eprint={1804.04811},
      archivePrefix={arXiv},
      primaryClass={cs.RO},
      url={https://arxiv.org/abs/1804.04811}, 
}

@misc{grandia2019mpc,
      title={{Feedback MPC for Torque-Controlled Legged Robots}}, 
      author={Ruben Grandia and Farbod Farshidian and René Ranftl and Marco Hutter},
      year={2019},
      eprint={1905.06144},
      archivePrefix={arXiv},
      primaryClass={cs.RO},
      url={https://arxiv.org/abs/1905.06144}, 
}

@misc{adabag2025toyota,
      title={{Differentiable Model Predictive Control on the GPU}}, 
      author={Emre Adabag and Marcus Greiff and John Subosits and Thomas Lew},
      year={2025},
      eprint={2510.06179},
      archivePrefix={arXiv},
      primaryClass={math.OC},
      url={https://arxiv.org/abs/2510.06179}, 
}

@article{chai2020mpc,
   title={{Adaptive and Efficient Model Predictive Control for Booster Reentry}},
   volume={43},
   ISSN={1533-3884},
   url={http://dx.doi.org/10.2514/1.G004925},
   DOI={10.2514/1.g004925},
   number={12},
   journal={Journal of Guidance, Control, and Dynamics},
   publisher={American Institute of Aeronautics and Astronautics (AIAA)},
   author={Chai, Joseph and Medagoda, Eran and Kayacan, Erkan},
   year={2020},
   month=dec,
   pages={2372–2382},
}

@article{pratama2024auto,
   title={{Non-linear model predictive control with single-shooting method for autonomous personal mobility vehicle}},
   volume={15},
   ISSN={2087-3379},
   url={http://dx.doi.org/10.55981/j.mev.2024.1105},
   DOI={10.55981/j.mev.2024.1105},
   number={2},
   journal={Journal of Mechatronics, Electrical Power, and Vehicular Technology},
   publisher={National Research and Innovation Agency},
   author={Pratama, Rakha Rahmadani and Baskoro, Catur Hilman Adritya Haryo Bhakti and Setiawan, Joga Dharma and Dewi, Dyah Kusuma and Paryanto, Paryanto and Ariyanto, Mochammad and Saputra, Roni Permana},
   year={2024},
   month=dec,
   pages={186–196}
}

@article{schwenzer2021mpc,
  author  = {Schwenzer, M. and Ay, M. and Bergs, T. and others},
  title   = {Review on model predictive control: an engineering perspective},
  journal = {International Journal of Advanced Manufacturing Technology},
  volume  = {117},
  pages   = {1327--1349},
  year    = {2021},
  doi     = {10.1007/s00170-021-07682-3}
}

@inproceedings{lazic2018hvac,
    title = {{Data Center Cooling using Model-predictive Control}},
    author = {Nevena Lazic and Tyler Lu and Craig Boutilier and MK Ryu and Eehern Jay Wong and Binz Roy and Greg Imwalle},
    year = {2018},
    URL	= {https://papers.nips.cc/paper/7638-data-center-cooling-using-model-predictive-control},
    booktitle = {Proceedings of the Thirty-second Conference on Neural Information Processing Systems (NeurIPS-18)},
    pages = {3818--3827},
    address	= {Montreal, QC},
}

@article{afram2017hvac,
    title = {{Artificial neural network (ANN) based model predictive control (MPC) and optimization of HVAC systems: A state of the art review and case study of a residential HVAC system}},
    journal = {Energy and Buildings},
    volume = {141},
    pages = {96-113},
    year = {2017},
    issn = {0378-7788},
    doi = {https://doi.org/10.1016/j.enbuild.2017.02.012},
    url = {https://www.sciencedirect.com/science/article/pii/S0378778816310799},
    author = {Abdul Afram and Farrokh Janabi-Sharifi and Alan S. Fung and Kaamran Raahemifar},
}

@article{benson2006direct,
author = {Benson, David and Huntington, Geoffrey and Thorvaldsen, Tom and Rao, Anil},
year = {2006},
month = {08},
pages = {1435-1440},
title = {{Direct Trajectory Optimization and Costate Estimation via an Orthogonal Collocation Method}},
volume = {29},
journal = {Journal of Guidance Control and Dynamics - J GUID CONTROL DYNAM},
doi = {10.2514/1.20478}
}

@article{raissi2019pinn,
title = {{Physics-informed neural networks: A deep learning framework for solving forward and inverse problems involving nonlinear partial differential equations}},
journal = {Journal of Computational Physics},
volume = {378},
pages = {686-707},
year = {2019},
issn = {0021-9991},
doi = {https://doi.org/10.1016/j.jcp.2018.10.045},
url = {https://www.sciencedirect.com/science/article/pii/S0021999118307125},
author = {M. Raissi and P. Perdikaris and G.E. Karniadakis},
}

@misc{zhong2025pinn,
      title={{A Physics-Informed Neural Networks-Based Model Predictive Control Framework for $SIR$ Epidemics}}, 
      author={Aiping Zhong and Baike She and Philip E. Paré},
      year={2025},
      eprint={2509.12226},
      archivePrefix={arXiv},
      primaryClass={cs.LG},
      url={https://arxiv.org/abs/2509.12226}, 
}

@article{regev2023hykkt,
    author = {Shaked Regev and Nai-Yuan Chiang and Eric Darve and Cosmin G. Petra and Michael A. Saunders and Kasia Świrydowicz and Slaven Peleš},
    title = {{HyKKT: a hybrid direct-iterative method for solving KKT linear systems}},
    journal = {Optimization Methods and Software},
    volume = {38},
    number = {2},
    pages = {332--355},
    year = {2023},
    publisher = {Taylor \& Francis},
    doi = {10.1080/10556788.2022.2124990},
    URL = {https://doi.org/10.1080/10556788.2022.2124990},
    eprint = {https://doi.org/10.1080/10556788.2022.2124990}
}

@article{golub2003hykkt,
    author = {Golub, Gene H. and Greif, Chen},
    title = {{On Solving Block-Structured Indefinite Linear Systems}},
    journal = {SIAM Journal on Scientific Computing},
    volume = {24},
    number = {6},
    pages = {2076-2092},
    year = {2003},
    doi = {10.1137/S1064827500375096},
    URL = {https://doi.org/10.1137/S1064827500375096},
    eprint = {https://doi.org/10.1137/S1064827500375096},
    }

@misc{pacaud2025kkt,
      title={{Condensed-space methods for nonlinear programming on GPUs}}, 
      author={François Pacaud and Sungho Shin and Alexis Montoison and Michel Schanen and Mihai Anitescu},
      year={2025},
      eprint={2405.14236},
      archivePrefix={arXiv},
      primaryClass={math.OC},
      url={https://arxiv.org/abs/2405.14236}, 
}

@InProceedings{chee2023node,
  title = 	 {{Learning-enhanced Nonlinear Model Predictive Control using Knowledge-based Neural Ordinary Differential Equations and Deep Ensembles}},
  author =       {Chee, Kong Yao and Hsieh, M. Ani and Matni, Nikolai},
  booktitle = 	 {Proceedings of The 5th Annual Learning for Dynamics and Control Conference},
  pages = 	 {1125--1137},
  year = 	 {2023},
  editor = 	 {Matni, Nikolai and Morari, Manfred and Pappas, George J.},
  volume = 	 {211},
  series = 	 {Proceedings of Machine Learning Research},
  month = 	 {15--16 Jun},
  publisher =    {PMLR},
  url = 	 {https://proceedings.mlr.press/v211/chee23a.html},
}

@article{schweidtmann2021neural,
author = {Schweidtmann, Artur M. and Esche, Erik and Fischer, Asja and Kloft, Marius and Repke, Jens-Uwe and Sager, Sebastian and Mitsos, Alexander},
title = {{Machine Learning in Chemical Engineering: A Perspective}},
journal = {Chemie Ingenieur Technik},
volume = {93},
number = {12},
pages = {2029-2039},
doi = {https://doi.org/10.1002/cite.202100083},
url = {https://onlinelibrary.wiley.com/doi/abs/10.1002/cite.202100083},
eprint = {https://onlinelibrary.wiley.com/doi/pdf/10.1002/cite.202100083},
year = {2021}
}

@article{zheng2023pinn,
  title   = {{Physics-Informed Online Machine Learning and Predictive Control of Nonlinear Processes with Parameter Uncertainty}},
  author  = {Zheng, Yingzhe and Wu, Zhe},
  journal = {Industrial \& Engineering Chemistry Research},
  volume  = {62},
  number  = {6},
  pages   = {2804--2818},
  year    = {2023},
  publisher = {American Chemical Society},
  doi     = {10.1021/acs.iecr.2c03691},
  url     = {https://doi.org/10.1021/acs.iecr.2c03691}
}

@Inbook{scott2024cholesky,
author={Scott, Jennifer and T{\r{u}}ma, Miroslav},
title={{Correction to: Algorithms for Sparse Linear Systems}},
bookTitle={Algorithms for Sparse Linear Systems},
year={2024},
publisher={Springer International Publishing},
address={Cham},
pages={C1--C2},
isbn={978-3-031-25820-6},
doi={10.1007/978-3-031-25820-6_12},
url={https://doi.org/10.1007/978-3-031-25820-6_12"}
}

@article{lin2005cholesky,
author = {Lin, Wen-Yang and Chen, Chuen-Liang},
title = {{On Optimal Reorderings of Sparse Matrices for Parallel Cholesky Factorizations}},
journal = {SIAM Journal on Matrix Analysis and Applications},
volume = {27},
number = {1},
pages = {24-45},
year = {2005},
doi = {10.1137/S0895479803386354},
URL = {https://doi.org/10.1137/S0895479803386354},
eprint = {https://doi.org/10.1137/S0895479803386354},
}

@article{forbes2015industry,
    title = {Model Predictive Control in Industry: Challenges and Opportunities},
    journal = {IFAC-PapersOnLine},
    volume = {48},
    number = {8},
    pages = {531-538},
    year = {2015},
    note = {9th IFAC Symposium on Advanced Control of Chemical Processes ADCHEM 2015},
    issn = {2405-8963},
    doi = {https://doi.org/10.1016/j.ifacol.2015.09.022},
    url = {https://www.sciencedirect.com/science/article/pii/S2405896315011039},
    author = {Michael G. Forbes and Rohit S. Patwardhan and Hamza Hamadah and R. Bhushan Gopaluni},
    }

@article{lubin2023jump,
  title={JuMP 1.0: Recent improvements to a modeling language for mathematical optimization},
  author={Lubin, Miles and Dowson, Oscar and Garcia, Joaquim Dias and Huchette, Joey and Legat, Beno{\^\i}t and Vielma, Juan Pablo},
  journal={Mathematical Programming Computation},
  volume={15},
  number={3},
  pages={581--589},
  year={2023},
  publisher={Springer}
}

\newpage
\appendix
\appendixpage
\begin{appendices}
\begin{figure}[!htb]
     \centering
     \begin{subfigure}[b]{0.48\textwidth}
         \centering
         \includegraphics[width=\textwidth]{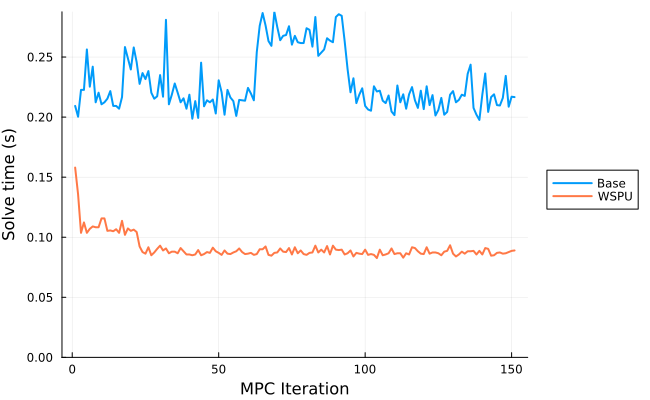}
         \caption{Distillation benchmark}
         \label{fig:distill-ipoptsolveits-combined}
     \end{subfigure}
     \hfill
     \begin{subfigure}[b]{0.48\textwidth}
         \centering
         \includegraphics[width=\textwidth]{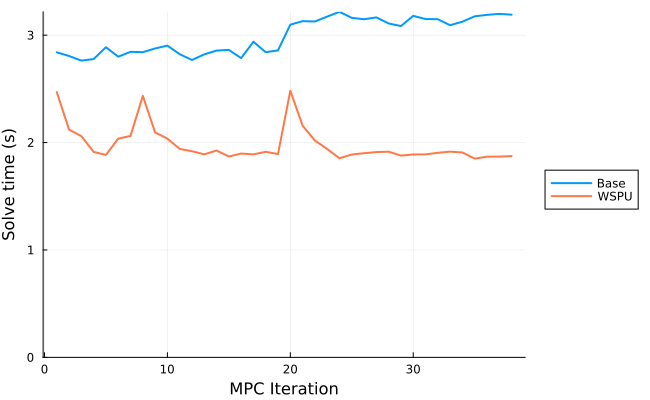}
         \caption{PDE plate benchmark}
         \label{fig:pde-ipoptsolveits-combined}
     \end{subfigure}
    \caption{Average JuMP + Reverse-mode AD + Ipopt solve times per NMPC iteration}
    \label{fig:benchmark-ipoptsolveits-combined}
\end{figure}

\begin{figure}[!htb]
     \centering
     \begin{subfigure}[b]{0.48\textwidth}
         \centering
         \includegraphics[width=\textwidth]{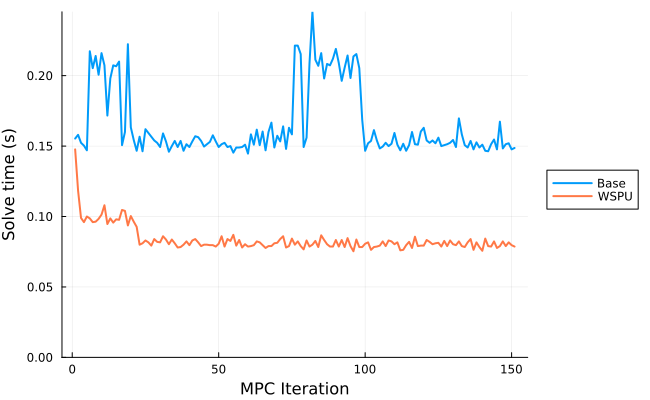}
         \caption{Distillation benchmark}
         \label{fig:distill-moisolveits-combined}
     \end{subfigure}
     \hfill
     \begin{subfigure}[b]{0.48\textwidth}
         \centering
         \includegraphics[width=\textwidth]{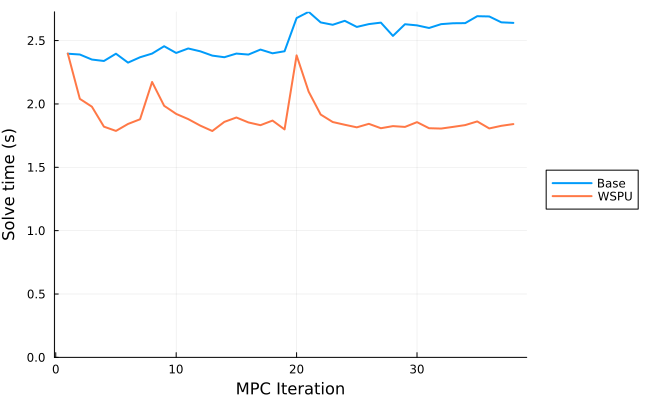}
         \caption{PDE plate benchmark}
         \label{fig:pde-moisolveits-combined}
     \end{subfigure}
    \caption{Average JuMP + Symbolic AD + Ipopt solve times per NMPC iteration}
    \label{fig:benchmark-moisolveits-combined}
\end{figure}

\begin{figure}[!htb]
     \centering
     \begin{subfigure}[b]{0.48\textwidth}
         \centering
         \includegraphics[width=\textwidth]{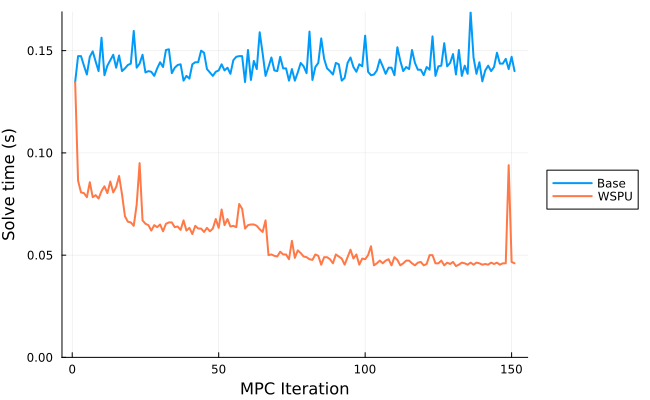}
         \caption{Distillation benchmark}
         \label{fig:distill-cpusolveits-combined}
     \end{subfigure}
     \hfill
     \begin{subfigure}[b]{0.48\textwidth}
         \centering
         \includegraphics[width=\textwidth]{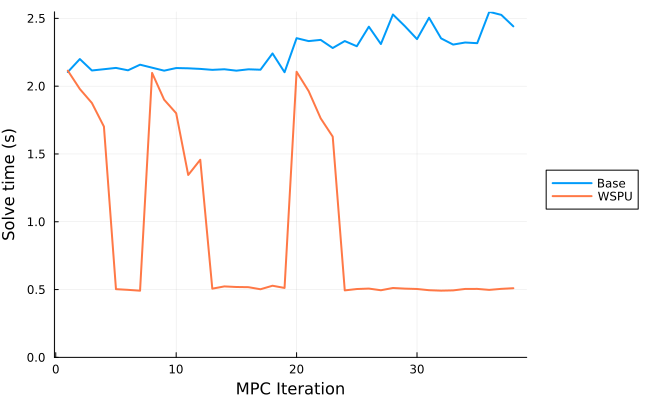}
         \caption{PDE plate benchmark}
         \label{fig:pde-cpusolveits-combined}
     \end{subfigure}
    \caption{Average ExaModelsCPU solve times per NMPC iteration}
    \label{fig:benchmark-cpusolveits-combined}
\end{figure}

\begin{table}[!htb]
\caption{Average Setup, Solve, Automatic Differentiation (AD), Linear Solver (Linear) and Total Times per NMPC Iteration for CPU and GPU  Configurations Under Different Warmstart (WS) and Parameter Update (PU) Settings for the Distillation Benchmark} \label{distill-bd}
\begin{adjustbox}{center=\textwidth}
\begin{tabular}{|c|c|c|c|c|c|c|}
    \hline
    \multicolumn{2}{|c|}{\multirow{2}{*}{\textbf{Configuration}}} & \multicolumn{5}{|c|}{\textbf{Times (s)}}\\\cline{3-7}
    \multicolumn{2}{|c|}{} & \textbf{Setup} & \textbf{Solve} & \textbf{AD} & \textbf{Linear} & \textbf{Iter. Total} \\
    \hline
    \multirow{4}{*}{\textbf{{\thead{JuMP +\\ Reverse-mode AD +\\ Ipopt}}}} & \textbf{Base} & 0.16 & 0.23 & 0.03 & 0.18 & 0.39 \\
    & \textbf{WS} & 0.16 & 0.18 & 0.02 & 0.14 & 0.34 \\
    & \textbf{PU} & 0.00 & 0.16 & 0.03 & 0.12 & 0.16 \\
    & \textbf{WS + PU} & 0.04 & 0.09 & 0.01 & 0.06 & 0.13 \\
    \hline
    \multirow{4}{*}{\textbf{{\thead{JuMP +\\ Symbolic AD +\\ Ipopt}}}} & \textbf{Base} & 0.15 & 0.17 & 0.01 & 0.13 & 0.32 \\
    & \textbf{WS} & 0.16 & 0.14 & 0.01 & 0.11 & 0.30 \\
    & \textbf{PU} & 0.00 & 0.20 & 0.01 & 0.17 & 0.20 \\
    & \textbf{WS + PU} & 0.04 & 0.08 & 0.01 & 0.07 & 0.12 \\
    \hline
    \multirow{4}{*}{\textbf{\thead{InfiniteExaModels +\\ NLPModelsIpopt}}} & \textbf{Base} & 0.01 & 0.14 & 0.00 & 0.12 & 0.15 \\
    & \textbf{WS} & 0.02 & 0.14 & 0.00 & 0.12 & 0.15 \\
    & \textbf{PU} & 0.00 & 0.14 & 0.00 & 0.12 & 0.14 \\
    & \textbf{WS + PU} & 0.00 & 0.06 & 0.00 & 0.05 & 0.06 \\
    \hline
    \multirow{4}{*}{\textbf{\thead{InfiniteExaModels +\\ MadNLP + cuDSS}}} & \textbf{Base} & 0.01 & 0.13 & 0.03 & 0.02 & 0.15 \\
    & \textbf{WS} & 0.13 & 0.13 & 0.03 & 0.01 & 0.25 \\
    & \textbf{PU} & 0.00 & 0.08 & 0.03 & 0.02 & 0.08 \\
    & \textbf{WS + PU} & 0.00 & 0.02 & 0.01 & 0.00 & 0.02 \\
    \hline
    \multirow{2}{*}{\textbf{\thead{OptimalControl +\\ ExaModels +\\ MadNLP + cuDSS}}} & 
    \rule[-2.5ex]{0pt}{5.0ex} \textbf{Base} & 0.04 & 0.09 & 0.02 & 0.02 & 0.13 \\
    & \rule[-2.0ex]{0pt}{3.0ex} \textbf{WS} & 0.04 & 0.06 & 0.01 & 0.01 & 0.10 \\
    \hline
    \textbf{MPCGPU} & \textbf{WS} & N/A & 0.02 & N/A & N/A & 0.02 \\
    \hline
\end{tabular}
\end{adjustbox}
\end{table}

\begin{table}[!htb]
\caption{Average Setup, Solve, Automatic Differentiation (AD), Linear Solver (Linear) and Total Times per NMPC Iteration for CPU and GPU  Configurations Under Different Warmstart (WS) and Parameter Update (PU) Settings for the PDE Plate Benchmark} \label{pde-bd}
\begin{adjustbox}{center=\textwidth}
\begin{tabular}{|c|c|c|c|c|c|c|}
    \hline
    \multicolumn{2}{|c|}{\multirow{2}{*}{\textbf{Configuration}}} & \multicolumn{5}{|c|}{\textbf{Times (s)}}\\\cline{3-7}
    \multicolumn{2}{|c|}{} & \textbf{Setup} & \textbf{Solve} & \textbf{AD} & \textbf{Linear} & \textbf{Iter. Total} \\
    \hline
    \multirow{4}{*}{\textbf{{\thead{JuMP +\\ Reverse-mode AD +\\ Ipopt}}}} & \textbf{Base} & 0.31 & 3.00 & 0.08 & 2.85 & 3.31 \\
    & \textbf{WS} & 0.31 & 2.00 & 0.06 & 1.88 & 2.31 \\
    & \textbf{PU} & 0.00 & 3.13 & 0.08 & 2.98 & 3.13 \\
    & \textbf{WS + PU} & 0.05 & 1.96 & 0.06 & 1.85 & 2.02 \\
    \hline
    \multirow{4}{*}{\textbf{{\thead{JuMP +\\ Symbolic AD +\\ Ipopt}}}} & \textbf{Base} & 0.34 & 2.52 & 0.07 & 2.38 & 2.86 \\
    & \textbf{WS} & 0.29 & 1.94 & 0.05 & 1.83 & 2.24 \\
    & \textbf{PU} & 0.00 & 2.52 & 0.07 & 2.38 & 2.52 \\
    & \textbf{WS + PU} & 0.06 & 1.88 & 0.05 & 1.77 & 1.94 \\
    \hline
    \multirow{4}{*}{\textbf{\thead{InfiniteExaModels +\\ NLPModelsIpopt}}} & \textbf{Base} & 0.01 & 2.26 & 0.02 & 2.18 & 2.27 \\
    & \textbf{WS} & 0.03 & 1.74 & 0.02 & 1.66 & 1.77 \\
    & \textbf{PU} & 0.00 & 2.22 & 0.02 & 2.13 & 2.22 \\
    & \textbf{WS + PU} & 0.00 & 0.93 & 0.01 & 0.89 & 0.93 \\
    \hline
    \multirow{4}{*}{\textbf{\thead{InfiniteExaModels +\\ MadNLP +\\ cuDSS}}} & \textbf{Base} & 0.01 & 0.96 & 0.02 & 0.68 & 0.97 \\
    & \textbf{WS} & 1.13& 0.32 & 0.01 & 0.15 & 1.45 \\
    & \textbf{PU} & 0.00 & 0.82 & 0.02 & 0.70 & 0.82 \\
    & \textbf{WS + PU} & 0.00 & 0.06 & 0.00 & 0.04 & 0.06 \\
    \hline
\end{tabular}
\end{adjustbox}
\end{table}

\end{appendices}
\end{document}